\documentclass[letterpaper,english]{aastex}
\usepackage[T1]{fontenc}
\usepackage{array}
\usepackage{longtable}
\usepackage{float}
\usepackage{textcomp}
\usepackage{amstext}
\usepackage{graphicx}
\usepackage{amssymb}
\usepackage{esint}

\makeatletter

\newcommand*\LyXbar{\rule[0.585ex]{1.2em}{0.25pt}}

\providecommand{\tabularnewline}{\\}

\slugcomment{}
\shorttitle{}
\shortauthors{}

\makeatother

\usepackage{babel}

\begin{document}

\title{A Mechanism for the Present-Day Creation of a New Class of Black
Holes}

\author{A. P. Hayes}

\affil{University of Maine, 120 Bennett Hall, Orono, Maine 04469-5709, USA}

\email{andrew.hayes@umit.maine.edu}

\author{N. F. Comins}

\affil{University of Maine, 120 Bennett Hall, Orono, Maine 04469-5709, USA}

\email{neil.comins@umit.maine.edu}
\begin{abstract}
In this first paper of a series on the formation and abundance of
substellar mass dwarf black holes (DBHs), we present a heuristic for
deducing the stability of non-rotating matter embedded in a medium
against collapse and the formation of a black hole. We demonstrate
the heuristic's accuracy for a family of spherical mass distributions
whose stability is known through other means. We also present the
applications of this heuristic that could be applied to data sets
of various types of simulations, including the possible formation
of DBHs in the expanding gases of extreme astrophysical phenomena
including Type Ia and Type II supernovae, hypernovae, and in the collision
of two compact objects. These papers will also explore the observational
and cosmological implications of DBHs, including estimates of the
total masses of these objects bound in galaxies and ejected into the
intergalactic medium. Applying our formalism to a Type II supernova
simulation, we have found regions in one data set that are within
a factor of three to four of both the density and mass necessary to
create a DBH.
\end{abstract}

\keywords{black hole physics \LyXbar{} dark matter \LyXbar{} dense matter \LyXbar{}
hydrodynamics \LyXbar{} instabilities \LyXbar{} supernovae:general}

\section{Introduction\label{sec:Introduction}}

Simulations reveal that supernovae are extremely chaotic, turbulent
events \citep{FW,FG} with {}``high speed fingers that emerge from
the core'' \citep{Burrows and Hayes-fingers} of the supernovae.
These simulated dynamics motivate us to investigate the plausibility
of small black holes, with masses less than 2 M$_{\odot}$, forming
in the ejecta of such events. The critical ingredient in such an investigation
is a criterion for determining whether an arbitrary volume of ejecta
is stable or whether it will undergo gravitational collapse. In non-relativistic
regimes, this is a relatively straightforward calculation of Jeans
instability \citep{Jeans}. In a general-relativistic regime, a general
theorem for stability becomes much less tractable; Chandrasekhar investigated
the problem and published several articles on it throughout his career,
declaring in \citet{Chandra Great Progress}:

\noindent \texttt{Great progress in the analysis of stellar structure
is not possible before we can answer the following question: Given
an enclosure containing electrons and atomic nuclei (total charge
zero), what happens if we go on compressing the material indefinitely?}

The conclusion of his investigation, decades later \citep{Chandra},
was a theorem describing the gravitational stability of a static,
non-rotating sphere in hydrostatic equilibrium in terms of general
relativistic field variables. Given the ongoing absence of a fully-general
theorem despite Chandrasekhar's early recognition of the importance
of the problem, we develop a limited, heuristic approach to finding
zones of gravitational collapse within more general mass distributions.

Harrison, Thorne, Wakano, and Wheeler (\citeyear{HTWW}, hereafter
HTWW), presented Chandrasekhar's stability theorem (Eq. 116 in HTWW)
in terms of physical quantities, unlike the original version, which
was stated in terms of general relativistic field variables. Chandrasekhar's
theorem proves the stability or instability against gravitational
collapse of a spherically symmetric, non-rotating, general-relativistic
distribution of matter in hydrostatic equilibrium bounded by the vacuum,
by calculating the squares of frequencies of eigenmodes. The theorem
is stated as a variational principle involving arbitrary spherical
perturbations, and the eigenmodes are found by successive minimizations
constrained by orthogonality to all previously found eigenmodes. If
any eigenmode is found to have a negative frequency squared, it will
grow exponentially rather than oscillate. Such growth implies that
the mass distribution is gravitationally unstable to small perturbations.
We expand this theorem into a heuristic that allows us to search for
regions of gravitational instability within non-spherical distributions
of matter, using the procedure described in Section \ref{sec:The-Chandrasekhar-Stability-Criterion}.

Allowing that there are volumes within the ejecta of supernovae, or
other extreme events, that become dwarf black holes (DBHs) by the
above criterion, it is then possible to calculate the mass spectrum
of the DBHs thereby ejected into the interstellar medium. In a subsequent
paper, we will convolve that spectrum with the supernova history of
a galaxy or cluster of galaxies to derive an estimate for the present-day
abundance of dwarf black holes produced by this process.

Our method can detect (within a simulation data set) a region of instability
that has the potential to form DBHs of mass less than 3 $M_{\odot}$
(highly dependent on which equation of state (EOS) is being used;
see Section \ref{sub:Stability Criterion Implementation}, particularly
Figure \ref{fig:M-vs-rho0}). There exists a phase change with respect
to different masses of DBH progenitors. Above this phase change, we
can arrive at a definitive diagnosis of stability or instability of
regions being considered as candidates for being DBH progenitors.
At the present time, below the phase change, we can only definitively
rule out instability. The particular value of the phase change, like
the maximum mass, depends sensitively and solely on the choice of
EOS. The density required to form a DBH increases rapidly as the total
mass of the DBH decreases, so we do not consider any DBH of mass less
than, say, gas giant planets, to be plausible, given the densities
currently found in simulations. The spatial resolution required for
definitive diagnosis of stability also increases rapidly with decreasing
DBH progenitor mass.

The phase change exists because of a feature intrinsic to the equations
governing the structure of spheres in hydrostatic equilibrium (to
be discussed in Sections \ref{sub:Smaller-Still-DBHs} and \ref{sub:Detecting-Higher-Order-Instabilities}).
Bodies above and below the phase change exhibit qualitatively different
behavior when compressed by exterior forces to the point of gravitational
collapse, and those below the phase change require a more complex
treatment. For these reasons and for the purpose of clarity, we will
refer to more massive, less dense, and more tractable DBHs above the
phase change as {}``Type I'' and those less massive, denser, less
tractable DBHs below the phase change as {}``Type II.''

Black holes of the Type I and plausible Type II mass ranges exhibit
negligible Hawking radiation \citep{Hawking} over the age of the
universe. These DBHs also have Schwarzschild radii under 5 km, and
thus present an astrophysically negligible cross section for direct,
photon-emitting mass capture and other interactions with gas and dust.
Dwarf black holes in the interstellar or intergalactic media, therefore,
are MACHOs (Massive Astrophysical Compact Halo Objects, with the possible
exception of {}``Halo''- to be discussed in a moment), in that they
are constructed from massive ensembles of ordinary particles and emit
very little or negligible electromagnetic radiation. Dwarf black holes,
however, may exhibit interesting longer range interactions, such as
seeding star formation while crossing HI clouds, generating turbulence
in the ISM, and scattering of matter disks surrounding stars, i.e.,
comet and asteroid disruption.

The possible contribution to galactic halo dark matter from any type
of MACHO in the somewhat substellar through 3 $M_{\odot}$ mass range
is poorly constrained by present observational projects searching
for MACHO dark matter \citep{Alcock96,Alcock98,FFG,TG}. These observational
programs execute large-scale photometric sweeps searching for microlensing
events within the Milky Way's halo and the Milky Way's immediate galactic
neighbors. These events are observable for compact lensing objects
of mass equal to the dwarf black holes we are examining. In addition,
\citet{FFG} found the best fit of mass for the MACHOs undergoing
microlensing events in their data set to be 0.5 $\text{M}_{\odot}$.

Supernova ejecta is expelled at very high speed, perhaps much greater
than the escape velocity of galaxies ($\sim$500 km s$^{\text{-1}}$
for the Milky Way, \citet{Carney and Latham-escapeV}), so it may
be that DBHs reside within the IGM rather than galactic halos. There,
DBHs would not be subject to the abundance constraints placed on MACHO
dark matter by microlensing surveys already completed. DBHs in the
IGM, unfortunately, will be extremely challenging to observe, probably
requiring large-scale pixel microlensing surveys. In Sections \ref{sub:Smaller-Still-DBHs}
and \ref{sub:Detecting-Higher-Order-Instabilities} we discuss the
possibility of detecting DBHs within simulation data sets.

Given their existence, DBHs will contribute to the mass of dark matter
in the universe. Our hypothesis, that black hole MAC(H)Os with negligible
Hawking radiation are being continuously created, predicts that the
contribution of DBHs to the dark matter content of the universe should
increase with time, and hence decrease with redshift.

\section{Hydrostatic Equations\label{sec:Hydrostatic-Equations}}

The spherically symmetric equations of hydrostatic equilibrium presented
below are derived in Misner, Thorne and Wheeler (\citeyear{MTW},
hereafter MTW). We follow their notation apart from re-including factors
of $c$ and $G$. All frame-dependent variables relate to rest frame
quantities.

In this section we are dealing with spherically symmetric ensembles
of matter, which are more mathematically accessible than the most
general distributions of matter, as pointed out by MTW, p. 603. In
spherical ensembles, the structure is determined by two coupled differential
equations, one for the mass enclosed within a given radius and one
for the pressure throughout the volume of the ensemble as a function
of radius. Further equations must then be added to close the system
of equations, as discussed in detail later in this section and Section
\ref{sub:Background}. MTW give the integral form of the following
equation for the radial mass distribution:

\begin{equation}
\frac{\mathrm{d}m(r)}{\mathrm{d}r}=\frac{4\pi}{c^{2}}r^{2}\rho(r)\label{eq:dmdr}\end{equation}
where $m(r)$ is the mass enclosed within $r$ (in units of mass),
and $\rho(r)$ is the mass-energy density (in units of energy per
volume). The simple form of this equation, familiar from its similarity
to its Newtonian analog, belies its relativistic complexity. The density,
$\rho(r),$ and the enclosed mass, $m(r)$, include energy content
as well as rest mass, and $r$ is the Schwarzschild radial coordinate
rather than the simple Newtonian radius. Henceforth we will abbreviate
{}``Schwarzschild radial coordinate'' as {}``radius'' but it should
always be construed in the general relativistic coordinate sense. 

The other equation of spherical structure that we need describes how
the pressure varies throughout the sphere. The general relativistic
correction to the Newtonian rate of change of pressure was first calculated
by \citet{OV}:

\begin{equation}
\frac{\mathrm{d}P(r)}{\mathrm{d}r}=-\frac{G}{c^{2}r^{2}}s(r)^{2}\left(\rho(r)+P(r)\right)\left(m(r)+\frac{4\pi}{c^{2}}r^{3}P(r)\right)\label{eq:dPdr}\end{equation}

where $P(r)$ is the pressure at a given radius, and $s(r)\equiv\left(1-\frac{2Gm(r)}{c^{2}r}\right)^{-\frac{1}{2}}$
is the factor appearing in the Schwarzschild solution to Einstein's
equations.

Equations \ref{eq:dmdr} and \ref{eq:dPdr} are fully general in the
sense that they can describe any fluid sphere in hydrostatic equilibrium.
These could be stable solutions ranging from a drop of water in zero
g to neutron stars, or solutions gravitationally unstable to perturbations.
The set of dependent variables in the system of Equations \ref{eq:dmdr}
and \ref{eq:dPdr} ($m(r)$, $P(r)$, and $\rho(r)$) has one more
member than we have equations and thus at least one more equation
must be added to close the system. This equation of state (EOS), which
is medium-dependent, relates pressure to density. We discuss the EOS
in greater detail in Section \ref{sec:The-Equation-of-State}.

We supplement the coupled structure equations, Equations \ref{eq:dmdr}
and \ref{eq:dPdr}, and the EOS, with two more equations that provide
significant insight into the dynamics of symmetric spheres even though
the former equations do not couple to them. The first supplemental
equation describes the number of baryons, $a(r)$, enclosed in a given
radius. From MTW, p. 606 (noting that their $e^{\Lambda}$ is equivalent
to our $s(r)$) :

\begin{equation}
a(r)=\int_{0}^{r}4\pi\bar{r}^{2}n(\bar{r})s(\bar{r})\,\mathrm{d}\bar{r}\end{equation}

Taking the derivative of $a$ with respect to $r$ gives

\begin{equation}
\frac{\mathrm{d}a(r)}{\mathrm{d}r}=4\pi r^{2}n(r)s(r)\label{eq:dAdr}\end{equation}
where $n(r)$ is the baryon number density as a function of $r$.
To calculate $a(r)$, our EOS must include $n$. The largest value
of $r$ where $n(r)>0$, denoted $R$, is the radius of the mass distribution.
Furthermore, we define $A\equiv a(R)$, the total baryon number within
$R$, and $M\equiv m(R)$, the total mass of the configuration as
measured by a distant observer. The insight provided by Equation \ref{eq:dAdr}
stems from $A$ being the only easily calculable physical quantity
that can show that two different configurations are merely rearrangements
of the same matter (modulo weak interactions changing the species
of baryons). For example, one Fe$^{\text{56}}$ nucleus has a specific
mass and radius while configurations of 56 H nuclei or 14 $\alpha$
particles will in general have different $R$s and, because of different
binding energies, different $M$s. All of the above will have an $A$
of 56, however.

Our second supplemental equation describes $\Phi(r)$, the general
relativistic generalization of the Newtonian gravitational potential.
From MTW, p. 604:

\begin{equation}
\frac{\mathrm{d}\Phi(r)}{\mathrm{d}r}=\frac{G}{r^{2}}s(r)^{2}\left(m(r)+\frac{4\pi}{c^{2}}r^{3}P(r)\right)\label{eq:dPhidr}\end{equation}

This equation is necessary if we want to perform any general relativistic
analysis of the dynamics of objects and spacetime in and around the
spheres we calculate. It allows us to, e.g., determine the spacetime
interval between two events and therefore whether and how they are
causally connected, or translate our Schwarzschild radial coordinate
into a proper radius, i.e. the distance that an object must actually
travel in its own reference frame to reach the origin.

\section{The Equation of State\label{sec:The-Equation-of-State}}

\subsection{Background\label{sub:Background}}

To solve the system of Equations \ref{eq:dmdr}, \ref{eq:dPdr}, and
optionally Equations \ref{eq:dAdr} and/or \ref{eq:dPhidr}, we must
relate pressure and density. In its most physically fundamental and
general form, this relation can be written as an expression of the
dependence of both $P$ and $\rho$ on thermodynamic state variables:
\begin{equation}
P=P(n,X)\label{eq:Peos}\end{equation}
\begin{equation}
\rho=\rho(n,X)\label{eq:rhoeos}\end{equation}
where $X$ is any other quantity or quantities on which the pressure-density
relations might depend, such as temperature, neutron-proton ratio,
or entropy per baryon. The variables $P$, $\rho,$ $n$ and/or $X$
may be subscripted to denote multiple particle species as appropriate
(MTW, p. 558). When necessary, the sums over particle species should
be used in Equations \ref{eq:dmdr}-\ref{eq:dPhidr}. Note that any
$X$ parameter must be controlled by additional, independent physics,
such as radiation transport, the assumption of a uniform temperature
distribution, or the enforcement of equilibrium with respect to beta
decay. That dependence is completely invisible to the structure equations,
however.

The pressure and mass-energy density must satisfy the local relativistic
formulation of the First Law of Thermodynamics (MTW, p. 559), i.e.\begin{equation}
\left(\frac{\mathrm{\partial}\rho}{\mathrm{\partial}n}\right)_{X}=\frac{P+\rho}{n}\equiv\mu\label{eq:mu}\end{equation}
where $\mu$ is the chemical potential, i.e. the energy, including
rest mass, necessary to insert one baryon into a small sample of the
material while holding all other thermodynamic quantities (denoted
here by $X$, similarly to Equations \ref{eq:Peos} and \ref{eq:rhoeos})
constant. We see that Equations \ref{eq:Peos}-\ref{eq:mu} overspecify
$P$, $\rho$, and $n$, so only two of these equations need be expressed
explicitly. If the EOS is specified in terms of Equations \ref{eq:Peos}
and \ref{eq:rhoeos}, they must be constructed so as to also obey
Equation \ref{eq:mu}. Pressure and mass-energy density may also appear
in a single equation expressed in terms of each other without any
reference to number density at all, though this precludes the calculation
of total baryon number of the sphere.

We introduce $\mu$ because it is useful in numerical simulations.
Overspecifying the thermodynamic characteristics of a sample of material
allows one to use whatever subset of variables will lend the best
resolution, least computational expense, and/or numeric stability
to calculations (e.g., \citet{LS}, p. 17). Using ordinary temperature
as the sole measure of {}``hotness'' of a material, for instance,
can lead to numerical instability in regions where the physics couples
only lightly to temperature such as in degenerate regions. The chemical
potential, in contrast, can continue to be a useful indicator of the
same information throughout the region of interest.

The {}``universal'' equation of state is in principle a measurable,
tabulatable quantity with a single {}``right answer'' for any given
composition and set of thermodynamic state variables that spans the
composition's degrees of freedom. However, the massive domain and
extreme complexity of physics that such an equation of state, describing
all objects everywhere, would encompass renders its formulation unfeasible.
We can treat only small patches of the EOS domain at a time. The region
of the equation of state domain of interest to this investigation
covers the extreme densities and pressures found in violent astrophysical
phenomena, and usually electrically-neutral compositions. (With that
in mind, we will discuss {}``baryons'' with the implicit assumption
of the presence of sufficient leptons as appropriate to the context.)
This region is far beyond the reach of present and perhaps all future
terrestrial experiments; it therefore must be investigated by theory
and the inversion of astronomical observations of extreme phenomena.
There remain many different candidates for the right EOS in this regime,
a very active area of research. We therefore present Table \ref{tab:Equations-of-State},
a compilation of the EOSs that have been of interest to our investigation.
They include several past and present candidates. Each EOS, and the
reason that it is of interest to our research, is discussed in detail
in the proceeding sections.

\begin{table}[h]
\begin{tabular}{|>{\centering}p{1.5in}|>{\centering}p{1.25in}|c|}
\hline 
EOS & Fig. Abbrev. & Reference\tabularnewline
\hline
\hline 
Oppenheimer-Volkoff & frudgas & \citet{OV}\tabularnewline
\hline 
Harrison-Wheeler & hwtblgas & \citet{HW}\tabularnewline
\hline 
Myers \& Swiatecki & myersgas & \citet{MS}\tabularnewline
\hline 
Douchin \& Haensel & dhgas & \citet{DH}\tabularnewline
\hline 
Lattimer \& Swesty & lseos, sk1eos, skaeos, skmeos & \citet{LS}\tabularnewline
\hline 
Non-relativistic, ultradegenerate gas & nrudgas & \citet{Kippenhahn Weigert}\tabularnewline
\hline
\end{tabular}\caption{\emph{Equations of State} We tabulate the various equations of state
used in this paper. We also denote how they are abbreviated when referenced
in Figures in this paper, and from what source they came. They are
discussed in more detail throughout the rest of this section.\label{tab:Equations-of-State}}
\end{table}

\subsection{Oppenheimer-Volkoff EOS\label{sub:Oppenheimer-Volkoff-EOS}}

The simplest EOS we consider assumes that pressure arises solely from
the momentum flux of neutrons obeying Fermi-Dirac statistics at absolute
zero, called a Fermi gas \citep{OV}. Relativistic effects are treated
fully. We include this EOS in this investigation because it was used
in the original \citet{OV} paper proposing the existence of neutron
stars. Our integration algorithm (to be discussed in detail in Section
\ref{sec:Solving-the-Structure-Equations}) reproduces within 1\%
their calculation of 0.71 solar masses for the Tolman-Oppenheimer-Volkoff
(TOV) limit \citep{OV} for the Oppenheimer-Volkoff (OV) EOS. The
value of the TOV limit will vary depending on which EOS is chosen.
Our faithful reproduction of the original, eponymous TOV limit gives
us confidence in the accuracy of our integration algorithm and its
implementation.

As an actual candidate for the real-world, high-density equation of
state, however, the OV EOS must be rejected because its TOV limit
is much less than the mass of observed neutron stars, for instance
the very heavy 1.97 M$_{\odot}$ neutron star observed by \citet{Heavy NS}.
The OV EOS' low TOV limit arises from the neglect of the strong force's
repulsive character at densities comparable to atomic nuclei \citep{DH};
such densities are also seen inside neutron stars and in astrophysical
explosions. Calculations assuming this EOS are abbreviated as frudgas
(Fully Relativistic, Ultra-Degenerate GAS) in figures to follow.

\subsection{Harrison-Wheeler EOS\label{sub:Harrison-Wheeler-EOS}}

The Harrison-Wheeler EOS \citep{HW} is numerically very similar to
the OV EOS of Section \ref{sub:Oppenheimer-Volkoff-EOS} in the regimes
most pertinent to neutron star bulk properties, but includes much
more physics in arriving at results and is additionally applicable
down to terrestrial matter densities. It is explicitly designed to
be the equation of state of matter at absolute zero and {}``catalyzed
to the endpoint of thermonuclear evolution'' (HTWW). In other words,
this EOS describes the ground states of ensembles of baryons at given
pressures, other than configurations containing singularities. The
Harrison-Wheeler EOS includes four separate regimes. They are, from
least to most dense, {}``Individualistic collections of {[}almost
entirely Fe$^{\text{56}}${]} atoms,'' (HTWW); {}``Mass of practically
pure Fe$^{\text{56}}$ held together primarily by chemical forces,''
(HTWW); a transitional regime where neutrons begin to increase in
abundance, forcing baryons to rearrange from Fe$^{\text{56}}$ to
Y$^{\text{122}}$ atoms; and, finally, a gas dominated by neutrons.
The Harrison-Wheeler EOS does not include the solid neutron crystal
phase found in the crust of neutron star models using modern equations
of state. We include the Harrison-Wheeler EOS because it is used in
HTWW for all their results, and we have made extensive comparisons
and citations to that reference. Calculations assuming this EOS are
abbreviated as hwtblgas in figures to follow.

\subsection{Myers \& Swiatecki EOS\label{sub:Myers-&-Swiatecki-EOS}}

In a pair of papers, Myers and Swiatecki (\citeyear{Other MS} and
\citeyear{MS}) formulated an EOS {}``based on the semiclassical
Thomas-Fermi approximation of two fermions per $h^{3}$ of phase space,
together with the introduction of a short-range (Yukawa) effective
interaction between the particles,'' \citep{MS}. Here, $h$ is Planck's
constant. They solved the Thomas-Fermi approximation for six adjustable
parameters and then found the values of these parameters that best
fit available empirical data, including corrections {}``For {[}nucleon
orbital{]} shell and even-odd {[}proton and neutron{]} effects and
for the congruence/Wigner energy,'' \citep{MS}, that the Thomas-Fermi
model cannot include. They further explain these corrections in \citet{Other MS}.
This EOS was of interest to us because it is expressed in closed form.
We found it useful to compare that with the more common tabulated
results, and also allows for unlimited extrapolation to high pressures
and densities. While not being necessarily reliable for physical predictions,
realizing these extreme extrapolations was necessary for illustrating
our discussion in Sections \ref{sub:Smaller-Still-DBHs} and \ref{sub:Detecting-Higher-Order-Instabilities}.
Calculations assuming this EOS are abbreviated as myersgas in figures
to follow.

\subsection{Douchin \& Haensel EOS\label{sub:Douchin-&-Haensel-EOS}}

The EOS obtained by \citet{DH} is calculated from the Skyrme Lyon
(SLy) potential \citep{Skyrme} approximated at absolute zero and
with nuclear and chemical interactions catalyzed to produce a ground
state composition. This EOS is published as a set of two tables, one
on either side of the phase transition between the solid neutron crystal
crust of a neutron star and its liquid interior. We included this
EOS to investigate the effects of including multiple phases of dense
matter on the properties of spherically symmetric ensembles. Calculations
assuming this EOS are abbreviated as DHgas in figures to follow.

\subsection{Lattimer \& Swesty EOS\label{sub:Lattimer-&-Swesty-EOS}}

\citet{LS} derived an EOS from the Lattimer, Lamb, Pethick, and Ravenhall
compressible liquid drop model \citep{LLPR} that includes photons,
electrons, and positrons approximated as ultrarelativistic, noninteracting
bosons or fermions, as appropriate; free nucleons; alpha particles
standing in for the mixture of light nuclei present; and heavier nuclei
represented as a single species with continuously variable composition.
Particles are assumed to be in equilibrium with respect to strong
and electromagnetic interactions but not necessarily in beta equilibrium.
Lattimer and Swesty demonstrate how the subset of configurations that
correspond to beta equilibrium can be selected from the larger domain.

This EOS was of interest to us because it includes both finite temperatures
and variable composition, so it simultaneously tests all features
of our integration routine discussed in Section \ref{sec:Solving-the-Structure-Equations}.
Several different nuclear interaction parameters are adjustable in
this EOS since they are not currently well-measured, and Lattimer
and Swesty have published four sets of tables of thermodynamic quantities
corresponding to four different choices of these parameters. Calculations
incorporating each of these choices are, respectively, abbreviated
as lseos, sk1eos, skaeos, and skmeos in figures to follow, after Lattimer
and Swesty's own notation. Sk1 is also used by \citet{FY2007}, so
it was required to make a valid comparison between our results and
Fryer's supernova simulation data as described in Section \ref{sec:The-Chandrasekhar-Stability-Criterion}.

\subsection{Non-relativistic Ultra-degenerate Gas EOS\label{sub:Non-relativistic-Ultra-degenerate-Gas-EOS}}

We included a non-relativistic EOS adapted from the detailed treatment
of a Fermi gas in \citet{Kippenhahn Weigert} as a check for consistency.
This equation of state assumes, like the Oppenheimer-Volkoff EOS,
that the pressure results solely from the momentum flux of neutrons
obeying the Pauli exclusion principle at absolute zero, but includes
no relativistic effects. The mass-energy density in the Non-relativistic
Ultra-degenerate Gas EOS is calculated only as the rest-mass density,
and the speed of light has no special significance. In particular,
kinetic energy is calculated solely as $\frac{1}{2}mv^{2}$ and high-energy
particle states achieved by very dense samples of matter may have
velocities greater than $c$. The inconsistencies inherent in this
equation of state, especially when used with General Relativistic
structure equations, produced some surprising and easily distinguished
artifacts when used to produce a spherically symmetric model, which
we discuss in Section \ref{sub:Stability Criterion Implementation}.
Calculations assuming this EOS are abbreviated as nrudgas (Non-Relativistic,
Ultra Degenerate GAS) in figures to follow.

\section{Solving the Structure Equations\label{sec:Solving-the-Structure-Equations}}

Equations \ref{eq:dmdr} and \ref{eq:dPdr} and any two of Equations
\ref{eq:Peos}, \ref{eq:rhoeos}, and \ref{eq:mu} can be solved together.
To do this numerically, we begin by selecting a central pressure as
a boundary condition and solving Equation \ref{eq:Peos} for $n$
in terms of $P$ in order to find the central number density. We then
use Equation \ref{eq:rhoeos} to find the mass-energy density. We
are interested in the gravitational potential, so we also use Equation
\ref{eq:dPhidr}. We complete our calculation by changing the gauge
of $\Phi(r)$ when we know its asymptotic value, but for now we begin
with an arbitrary central boundary condition of $\Phi(0)=0$. All
of our EOSs provide $n$, so we also calculate $a(r)$ from Equation
\ref{eq:dAdr}. The structure equations then provide the radial derivatives,
so we integrate $m(r),$ $P(r),$ and $a(r)$ outwards using a numeric
algorithm (discussed in more detail later in this section). Physical
consistency between the integrations of Equations \ref{eq:dmdr} and
\ref{eq:dPdr} is maintained by enforcing the chosen EOS at each radial
point. If the EOS is not formulated as $\rho(P)$ in the literature,
we solve the EOS analytically if possible or numerically if necessary
in order to achieve the functional relationship of $\rho$ as a function
of $P$.

Our integration stops when $P(r)$ reaches or overshoots zero. Recall
that we set $R\equiv r$, the value of the radial coordinate where
the integration ends, $M\equiv m(R)$, the total mass of the sphere
as measured by a distant observer, and $A\equiv a(R)$ , the total
baryon number of the sphere.

At the end of our integration, it is necessary to change the gauge
of $\Phi(r)$ so that $\Phi(\infty)=0$. This allows us to make valid
comparisons to what the customary relativistic {}``Observer at Infinity''
observes. We determine the gauge by first noting the form that Equation
\ref{eq:dPhidr} takes in the exterior, vaccuum region $r>R$. Substituting
our definition of $s(r)$, and the values of $m(r)$ and $P(r)$ in
the exterior region into Equation \ref{eq:dPhidr}, we have

\begin{equation}
\frac{\mathrm{d}\Phi(r)}{\mathrm{d}r}=\frac{GM}{r^{2}-2GMr/c^{2}}\;\left(\text{for }r\geq R\right)\end{equation}

The analytic solution in our desired gauge ($\Phi(\infty)=0$) is
$\Phi(r)=\left(c^{2}/2\right)\ln\left(1-2GM/c^{2}r\right)$. We then
add a constant to the interior, numerical solution so that it matches
our exterior, analytic solution at $r=R$. Our construction guarantees
that $\mathrm{d}\Phi(r)/\mathrm{d}r$ is continuous at the junction
and therefore $\Phi(r)$ is smooth.

We use an Adams-Bashforth integration algorithm \citep{BA}, keeping
terms to fourth order in the following calculations. A uniformly-gridded
independent variable, in our case the Schwarzschild radial coordinate,
is characteristic of this algorithm. We have found that fourth-order
integration with a radial step size of 1 m reproduces within 1\% models
calculated with a higher order algorithm or step size an order of
magnitude smaller.

\section{The Chandrasekhar Stability Criterion\label{sec:The-Chandrasekhar-Stability-Criterion}}

\subsection{Theory\label{sub:Stability Criterion Theory}}

\citet{Chandra} developed a variational principle that calculates
the frequencies squared of eigenmodes of a given vaccuum-bounded static
sphere in hydrostatic equilibrium. The term {}``static sphere in
hydrostatic equilibrium'' includes initially unperturbed configurations
whose hydrostatic equilibrium is unstable. The original version of
Chandrasekhar's frequency mode theorem \citep{Chandra} for spherical
mass distributions in a vaccuum included difficult-to-characterize
variables from the general relativistic field equations. HTWW use
more physically intuitive variables for all instances of the field
equation variables, here in non-geometrized units:

\begin{equation}
\omega^{2}=\frac{\int_{0}^{R}J(r)\,\mathrm{d}r}{\int_{0}^{R}K(r)\,\mathrm{d}r}\label{eq:omega2}\end{equation}
where

\begin{eqnarray*}
J(r) & = & c^{6}\mu_{s}^{2}s(R)r^{-2}n(r)^{2}\mu(r)^{-3}\frac{\mathrm{d}\mu(n(r))}{\mathrm{d}n}s(r)\left[\frac{\mathrm{d}}{\mathrm{d}r}\left(r^{2}\mu(r)q(r)\right)\right]^{2}\\
 &  & -4c^{4}\mu_{s}^{2}s(R)Gn(r)s(r)^{3}\left(2\pi c^{-2}r^{2}P(r)+\frac{m(r)}{r}\right)q(r)^{2}\\
 &  & -c^{2}\mu_{s}^{2}s(R)G^{2}n(r)s(r)^{5}(4\pi c^{-2}r^{2}P(r)+\frac{m(r)}{r})^{2}q(r)^{2}\\
\end{eqnarray*}

\[
K(r)=s(r)^{3}n(r)\mu(r)^{2}r^{2}q(r)^{2}\]

\noindent where $\mu_{s}$ is 1/56 the mass of one free Fe$^{\text{56}}$
atom, $s(R)$ is the Schwarzschild factor at the surface of the sphere,
and $q(r)$ is the displacement from equilibrium as a function of
$r$ for the eigenmode in question.

If we were to find and sort in ascending order the squares of the
eigenfrequencies of every eigenmode, we would then need consider only
the fundamental mode: if the square of its eigenfrequency is positive,
corresponding to an oscillatory mode, all higher eigenmodes must also
be oscillatory. In contrast, if the square of the fundamental eigenmode's
frequency is negative, then that mode is unstable and grows exponentially.
The existence of even a single growth eigenmode demonstrates that
random perturbations would grow as well, making the sphere unstable
to gravitational collapse. To prove instability, we must therefore
minimize Equation \ref{eq:omega2}, guaranteeing that we have found
the fundamental eigenmode. This is nominally a calculus of variations
problem, very expensive to compute numerically. We can reduce the
complexity of the problem significantly, however, through the following
analysis:

We note that the denominator of Equation \ref{eq:omega2} is positive
definite for nontrivial spheres and vibrational modes. Since we are
not interested in the actual frequency but only the sign of the fundamental,
and have no interest in higher modes, we can henceforth ignore the
denominator and focus our attention solely on the numerator:\begin{eqnarray}
\omega^{2} & \propto & \int_{0}^{R}r^{-2}n(r)^{2}\mu(r)^{-3}\frac{\mathrm{d}\mu(n(r))}{\mathrm{d}n}s(r)\left[\frac{\mathrm{d}}{\mathrm{d}r}\left(r^{2}\mu(r)q(r)\right)\right]^{2}\,\mathrm{d}r\nonumber \\
 &  & -\int_{0}^{R}4Gc^{-2}n(r)s(r)^{3}\left(2\pi c^{-2}r^{2}P(r)+\frac{m(r)}{r}\right)q(r)^{2}\,\mathrm{d}r\\
 &  & -\int_{0}^{R}G^{2}c^{-4}n(r)s(r)^{5}(4\pi c^{-2}r^{2}P(r)+\frac{m(r)}{r})^{2}q(r)^{2}\,\mathrm{d}r\nonumber \\
\nonumber \end{eqnarray}

We can achieve greater insight into $\omega^{2}$ by temporarily generalizing
each $q(r)^{2}$ into $q_{1}(r)q_{2}(r)$. This separation does not
change the mathematical content of the equation.

\begin{eqnarray}
\omega^{2} & \propto & \int_{0}^{R}r^{-2}n(r)^{2}\mu(r)^{-3}\frac{\mathrm{d}\mu(n(r)}{\mathrm{d}n}s(r)\left[\frac{\mathrm{d}}{\mathrm{d}r}\left(r^{2}\mu(r)q_{1}(r)\right)\right]\left[\frac{\mathrm{d}}{\mathrm{d}r}\left(r^{2}\mu(r)q_{2}(r)\right)\right]\,\mathrm{d}r\nonumber \\
 &  & -\int_{0}^{R}4Gc^{-2}n(r)s(r)^{3}\left(2\pi c^{-2}r^{2}P(r)+\frac{m(r)}{r}\right)q_{1}(r)q{}_{2}(r)\,\mathrm{d}r\\
 &  & -\int_{0}^{R}G^{2}c^{-4}n(r)s(r)^{5}(4\pi c^{-2}r^{2}P(r)+\frac{m(r)}{r})^{2}q_{1}(r)q_{2}(r)\,\mathrm{d}r\nonumber \\
\nonumber \end{eqnarray}

We see that each integral takes the form of an inner product over
the vector space of real functions. We will express those inner products
as a single linear operator $\widehat{M}$, as such:

\begin{equation}
\omega^{2}\propto<q_{1}|\widehat{M}|q_{2}>\label{eq:omega2bracket}\end{equation}

\noindent where \begin{eqnarray}
\widehat{M} & = & r^{-2}n(r)^{2}\mu(r)^{-3}\frac{\mathrm{d}\mu(n(r))}{\mathrm{d}n}s(r)\left[\frac{\mathrm{d}}{\mathrm{d}r}\left(r^{2}\mu(r)\Leftarrow\right)\right]\left[\frac{\mathrm{d}}{\mathrm{d}r}\left(r^{2}\mu(r)\Rightarrow\right)\right]\nonumber \\
 &  & -4Gc^{-2}n(r)s(r)^{3}\left(2\pi c^{-2}r^{2}P(r)+\frac{m(r)}{r}\right)\\
 &  & -G^{2}c^{-4}n(r)s(r)^{5}(4\pi c^{-2}r^{2}P(r)+\frac{m(r)}{r})^{2},\nonumber \\
\nonumber \end{eqnarray}
 a $\Leftarrow$ indicates that $\widehat{M}$ acts on the bra vector
in that location, a $\Rightarrow$ similarly indicates that $\widehat{M}$
acts on the ket vector, and the inner product operation is taken to
be an integral from $0$ to $R$ over $\mathrm{d}r$. Note that we
can use the chain rule and integration by parts to express $\widehat{M}$
in the more familiar fashion of an operator acting only on the ket
vector. 

The expression of $\omega^{2}$ in the form of Equation \ref{eq:omega2bracket}
allows us to use the tools of linear algebra to achieve the aforementioned
reduction in the complexity of our problem via the following method.

Let $\left\{ \mathbf{v}{}_{j}\right\} $ be a complete, orthonormal
eigenbasis for $\widehat{M}$ with corresponding eigenvalues $\lambda_{j}$,
where $\widehat{M}\mathbf{v}{}_{j}=\lambda_{j}\mathbf{v}{}_{j}$.
Any vector $\mathbf{q}$ can be expressed as a linear combination,
$\mathbf{q}=\sum_{j}a_{j}\mathbf{v}{}_{j}$. We note by inspection
that 

\begin{equation}
<q|\widehat{M}|q>=\left(\sum_{j}a_{j}\mathbf{v}{}_{j}\right)\cdot\widehat{M}\left(\sum_{k}a_{k}\mathbf{v}{}_{k}\right)=\left(\sum_{j}a_{j}\mathbf{v}{}_{j}\right)\cdot\left(\sum_{k}a_{k}\lambda_{k}\mathbf{v}_{k}\right)=\sum_{k}a_{k}^{2}\lambda_{k}\end{equation}

For there to exist a $\mathbf{q}$ that yields a negative value for
$<q|\widehat{M}|q>$ , at least one $\lambda_{j}$ must be negative.
Our minimization problem is therefore reduced to finding the sign
of the lowest eigenvalue of the operator $\widehat{M}$. Unfortunately,
$\widehat{M}$ acts on an infinite-dimensional vector space, so to
proceed farther we must approximate $\widehat{M}$ as a matrix over
a finite-dimensional subspace with basis to be chosen in a moment.
We call our approximation $\widehat{M_{n}}$, where $n$ is the number
of basis vectors we include. Because we integrated Equations \ref{eq:dmdr}-\ref{eq:dPhidr}
numerically and thus have no analytic form for the physical variables
present in Equation \ref{eq:omega2}, we can not state \emph{a priori}
a closed form for the eigenspectrum of the $\widehat{M}$ operator
for a given sphere. Therefore, we must express displacements $q(r)$
in terms of an arbitrary basis, but which one? HTWW analyzed the general
case of the fundamental mode, which is a simultaneous expansion and/or
contraction of the entire star with a central node, as demanded by
symmetry. (If that mode is oscillatory, it will do one and then the
other. If it is a growth mode, it will do only one, in a runaway process.)
Let us select, then, the simplest possible basis, i.e. the monomials
$r^{i}$, where the exponent $i$ runs from 1 to $n$. We exclude
the negative powers of $r$ and $r^{0}$ because they violate our
symmetry condition of a central node. We will determine the lowest
eigenvalue in terms of this basis.

\subsection{Computational Implementation and Verification\label{sub:Stability Criterion Implementation}}

\begin{figure}[H]
\includegraphics{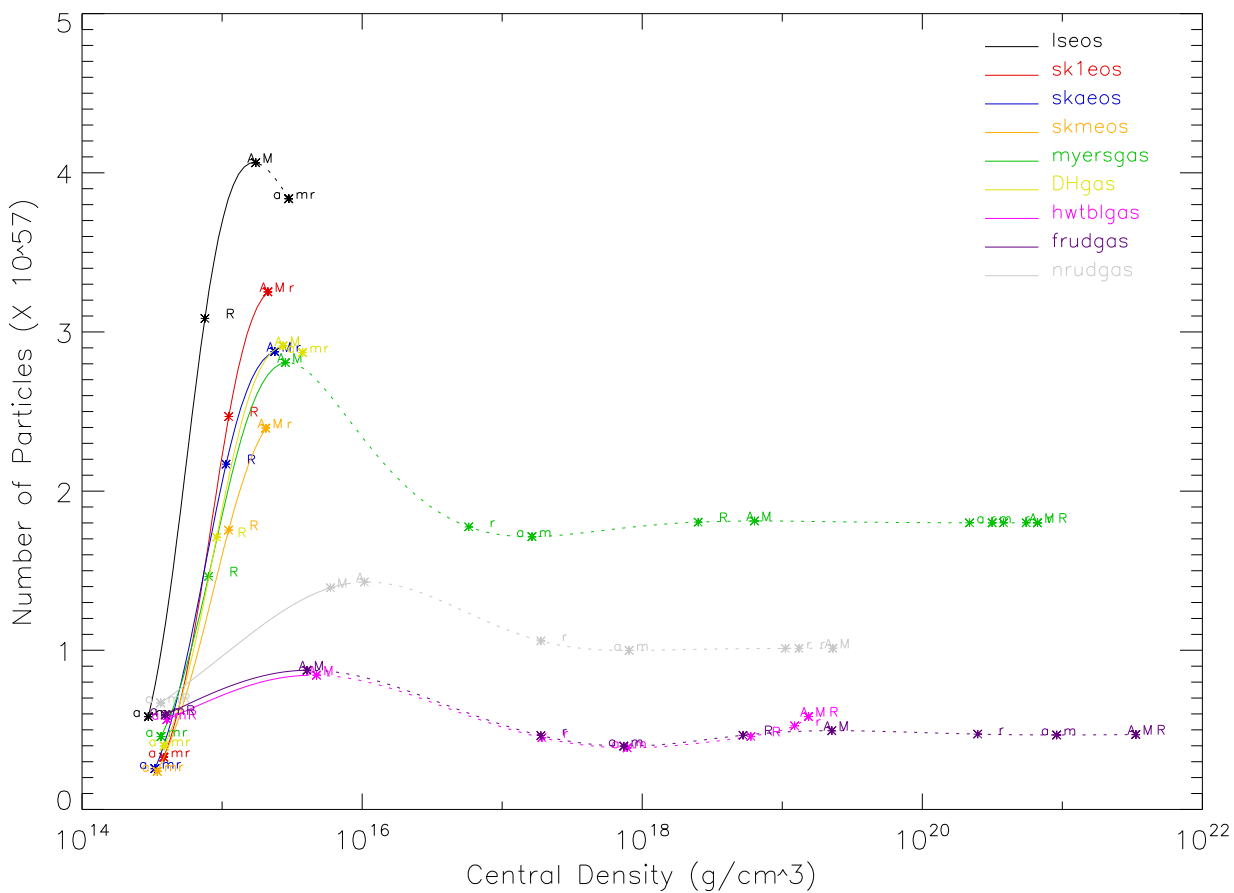}

\caption{\emph{Total Baryon number, A, vs Central Density, $\rho{}_{\text{0}}$,
of a self-gravitating sphere in hydrostatic equilibrium.} These curves
represent continuous sequences of the final results of integrations
of Equation \ref{eq:dAdr} using different central pressures as boundary
conditions for the system of differential equations of structure.
Each different line represents the choice of a different EOS. The
asterisked points on the curves represent configurations corresponding
to critical points of various quantities. The capital letters A, M,
and R indicate local maxima of baryon number, mass, and radius, respectively,
vs. $\rho{}_{\text{0}}$. Lower-case letters represent analogous local
minima. The curves are drawn solid where our implementation of the
Chandrasekhar stability theorem indicates that the solution corresponds
to stable hydrostatic equilibrium. The curves are drawn as dotted
lines where it indicates unstable hydrostatic equilibrium. HTWW showed
that solutions lying to the left of the maximum on any particular
curve shown here are stable, while solution lying to the right are
unstable. Note the close agreement between these two independent methods
of calculating stability. \label{fig:A-vs-rho0}}
\end{figure}

We select $n=20$, the largest possible $n$ for this basis that reliably
does not overflow 64-bit floating point variables. We expect that
such a large basis should be more than sufficient to accurately model
the true fundamental eigenmode, and indeed our results as shown in
Figure \ref{fig:A-vs-rho0} and discussed later in this section show
that to be the case. We found, in our calculations using the $r^{i}$
basis, that upper-left matrix elements of $M_{n}$, corresponding
to small powers of $r$, are many orders of magnitude smaller than
lower-right matrix elements. Having tabulated the matrix elements
of $M_{n}$ for a given sphere, we then use the built-in eigenvalue
function of ITTVIS' IDL language to determine the lowest value and
therefore the stability of the sphere.

Figure \ref{fig:A-vs-rho0} is a graph of the total baryon number
vs. central density (converted to mass units) of spheres in hydrostatic
equilibrium, calculated according to our previously discussed procedure,
for the EOSs presented in Section \ref{sec:The-Equation-of-State}.
Each curve represents a family of spheres of different central densities,
for a single equation of state. We draw the curves as solid lines
where the Chandrasekhar theorem-based stability criterion indicates
that the sphere is stable (i.e. $\omega^{2}>0$), while we draw the
curves as dotted lines where the criterion indicates instability (i.e.
$\omega^{2}<0$).

We can also interpret the stable (solid-line) portion of each line
as the construction, by slow mass accretion, of an object consistently
obeying a single EOS, to the point of collapse. As long as material
of the appropriate composition and temperature is added with negligible
bulk kinetic energy at a rate negligible within a hydrodynamic timescale,
a single object will move along a given line from left to right.

To continue moving the so-constructed object to the right, into and
through the unstable portion, requires a different action. We see
that the total mass of the object must decrease for a finite span,
even as the central density increases. This means simply that we have
reached the maximum possible mass that our chosen EOS can support,
and to achieve any greater compression in the core of the object we
must remove a quantity of matter while increasing that compression,
or else the object will immediately collapse. (Note that this must
be accomplished while somehow avoiding the introduction of the slightest
hydrodynamic perturbation not orthogonal to an overall compression.
No easy feat!) This technique will carry us to the first minimum of
an EOS's curve in Figure \ref{fig:A-vs-rho0} after the first maximum.
That family of solutions to the structure equation gives rise to Type
I DBHs. Material can then be added as before, which will bring us
to the second maximum, at a price of having to employ a still-more-sophisticated
technique to avoid introducing catastrophic perturbations. However,
as one might expect we can only return a very small fraction of the
removed matter before reaching that second maximum. We must then remove
material once again, and so on ad infinitum, in a strongly damped
oscillation of baryon number with respect to central density. All
of the configurations resulting from the latter manipulations correspond
to the progenitors of Type II DBHs, though not as simple identifications,
as with the Type I solutions.

\begin{figure}[H]
\includegraphics{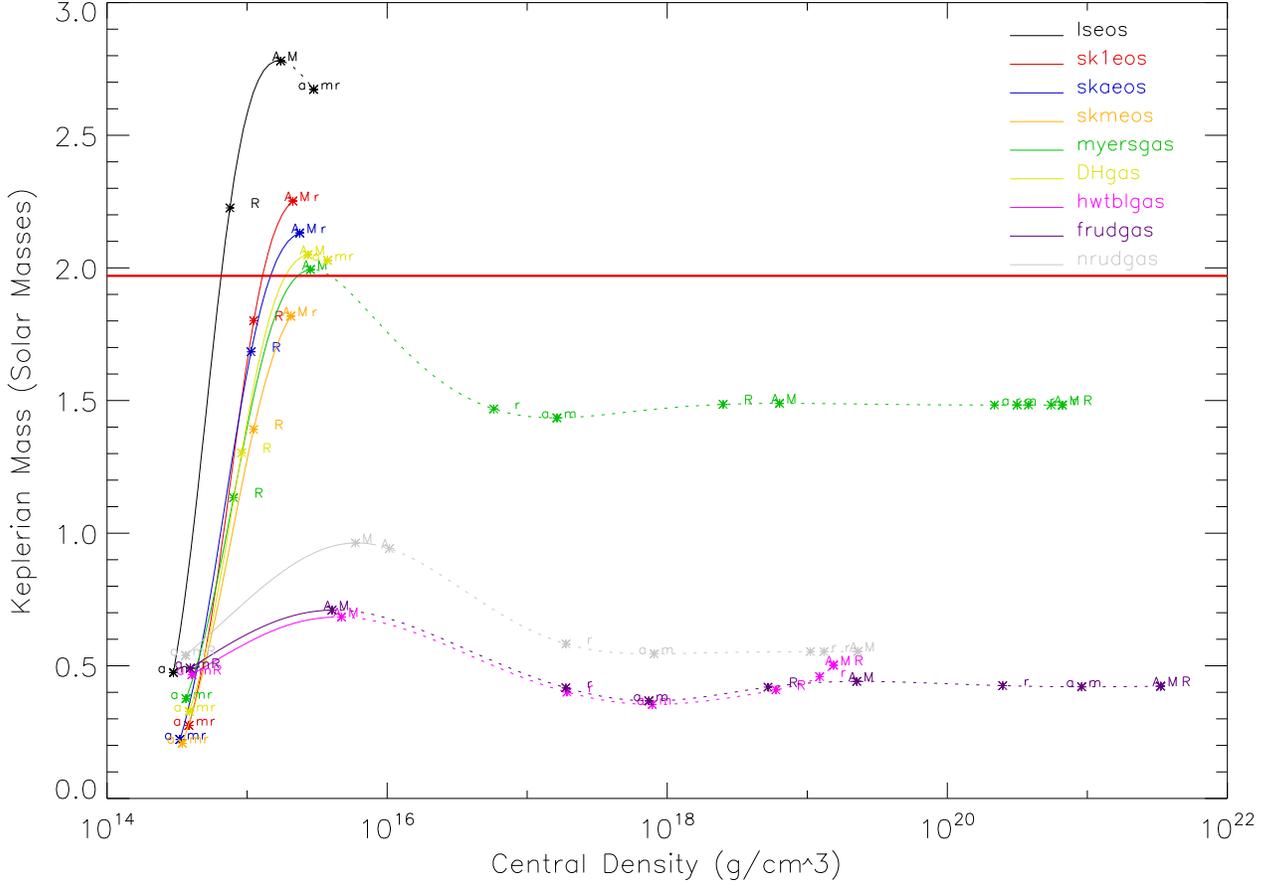}

\caption{\emph{Keplerian Mass, M, vs. Central Density, $\rho{}_{\text{0}}$,
of a self-gravitating sphere in hydrostatic equilibrium.} These curves
are drawn with the same conventions as those in Figure \ref{fig:A-vs-rho0}.
This graph shows us the total mass of a sphere in hydrostatic equilbrium
as measured by a very distant observer. The maximum of the \emph{frudgas}
curve (Fully Relativistic, UltraDegenerate GAS, the same EOS used
by \citet{OV}), demonstrates that we have accurately reproduced the
original, eponymous prediction of the TOV limit of 0.71 M$_{\odot}$.
The bold, red line indicates the mass of the neutron star observed
by \citet{Heavy NS}. Modulo corrections due to rotation, any EOS
whose maximum falls below this line must be rejected as incorrect.
\label{fig:M-vs-rho0}}
\end{figure}

Figure \ref{fig:M-vs-rho0} is a similar graph of total mass as measured
by an observer at infinity (the {}``Keplerian mass'') vs. central
density. It shows some subtle but critical differences from Figure
\ref{fig:M-vs-rho0} that we will elaborate with the next figure.
We include Figure \ref{fig:M-vs-rho0} because the Keplerian mass
is useful for comparisons to observations and the TOV limit for a
given EOS may be found directly as the maximum of that EOS's curve.
The bold, red line indicates the mass of the neutron star observed
by \citet{Heavy NS}. Although some corrections beyond the scope of
this work must be made for rotation, any EOS whose TOV limit falls
below this line must be rejected as incorrect.

We demonstrate the accuracy of our implementation of the stability
criterion by comparing it to an independent method of calculating
stability. This latter method (developed by HTWW, p. 50) is based
on a group property of a family of spheres calculated with the same
equation of state, but differing central densities. HTWW demonstrated
that spheres with lower central densities, i.e. left of the baryon
number maxima, denoted A in Figure \ref{fig:A-vs-rho0}, are in \emph{stable}
equilibrium and correspond to physical neutron stars, while those
at higher central density to the right of the mass maxima are unstable.
The close agreement between the family stability property (left of
the mass maximum is stable, right is unstable) and the single-sphere,
independent calculation of the Chandrasekhar stability criterion (whether
the curve is solid or dashed) gives us great confidence in our interpretation
and implementation of the latter.

\subsection{The Fate of an Unstable Equilibrium Configuration\label{sub:The-Fate-of-unstable}}

We know that unstable objects precisely at the TOV limit will collapse
gravitationally rather than disassociate because no stable equilibrium
configuration exists for that number of baryons, and no catastrophically
exothermic physics comes into play with our EOSs in this regime. The
fate of an unstable equilibrium configuration with central density
greater than the TOV-limit (located to the right of the first maximum
on each curve shown in Figures \ref{fig:A-vs-rho0} and \ref{fig:M-vs-rho0}),
however, is not predetermined.

\begin{figure}[H]
\includegraphics{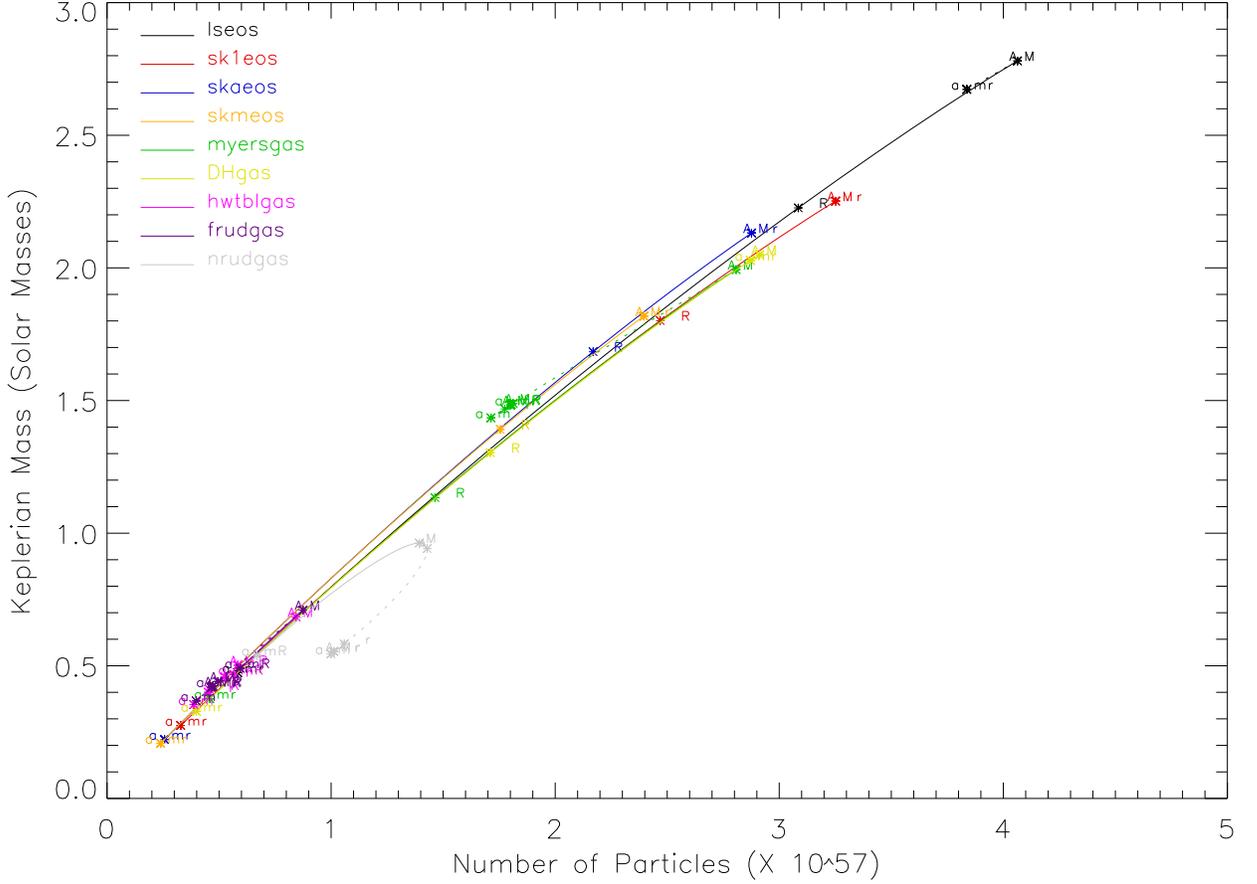}

\caption{\emph{Keplerian Mass, M, vs Total Baryon Number, A, of a self-gravitating
sphere in hydrostatic equilibrium}. These curves are drawn with the
same conventions as those in Figure \ref{fig:A-vs-rho0}. The vertical
distance between the stable equilibrium branch (solid) and unstable
equilibrium branch (dashed) of a given EOS for a given number of baryons
is the mass equivalent of the PV work necessary to compress the stable
equilibrium configuration to the verge of collapse. The non-relativistic
EOS included as a test for consistency (nrudgas, see Section \ref{sub:Non-relativistic-Ultra-degenerate-Gas-EOS})
is detected immediately as the one curve that takes a {}``wrong turn,''
as its mass calculation includes rest mass and binding energy but
assumes kinetic energy is negligible. \label{fig:M-vs-A}}
\end{figure}

Let us consider an informative merging of the previous two figures
in order to make evident the differences between the two. The results
are shown in Figure \ref{fig:M-vs-A}, where we have plotted the Keplerian
mass of objects in equilibrium versus their baryon number, i.e., correlated
values of the dependent variables of Figures \ref{fig:A-vs-rho0}
and \ref{fig:M-vs-rho0}. In addition to matching different variable
values for particular configurations as above, using Equation \ref{eq:dAdr}
to calculate baryon number also allows us to link different configurations
that correspond to rearrangements of the same amount of matter. Those
relationships would otherwise be impossible to detect.

Note that the non-relativistic EOS (nrudgas, See Section \ref{sub:Non-relativistic-Ultra-degenerate-Gas-EOS})
curve in Figure \ref{fig:M-vs-A} shows two unique deviations from
the behavior exhibited by the properly relativistic curves. The first
deviation is that it takes a {}``wrong turn''- the unstable branch
of the curve lies below the stable branch. This defect arises from
the neglect of kinetic energy, which is significantly greater in an
unstable configuration than the stable configuration of the same number
of baryons, in the calculation of the Keplerian mass. The second defect
is that while all the relativistic EOSs come to sharp points at the
upper right, the non-relativistic EOS is rounded. The discontinuity
of the parametric graph indicates that both the Keplerian mass and
baryon number are stationary with respect to their underlying independent
parameter, central pressure. The non-relativistic EOS' rounded end
indicates a defect in the calculation of the chemical potential, also
traceable to the neglect of kinetic energy.

\begin{figure}[H]
\includegraphics{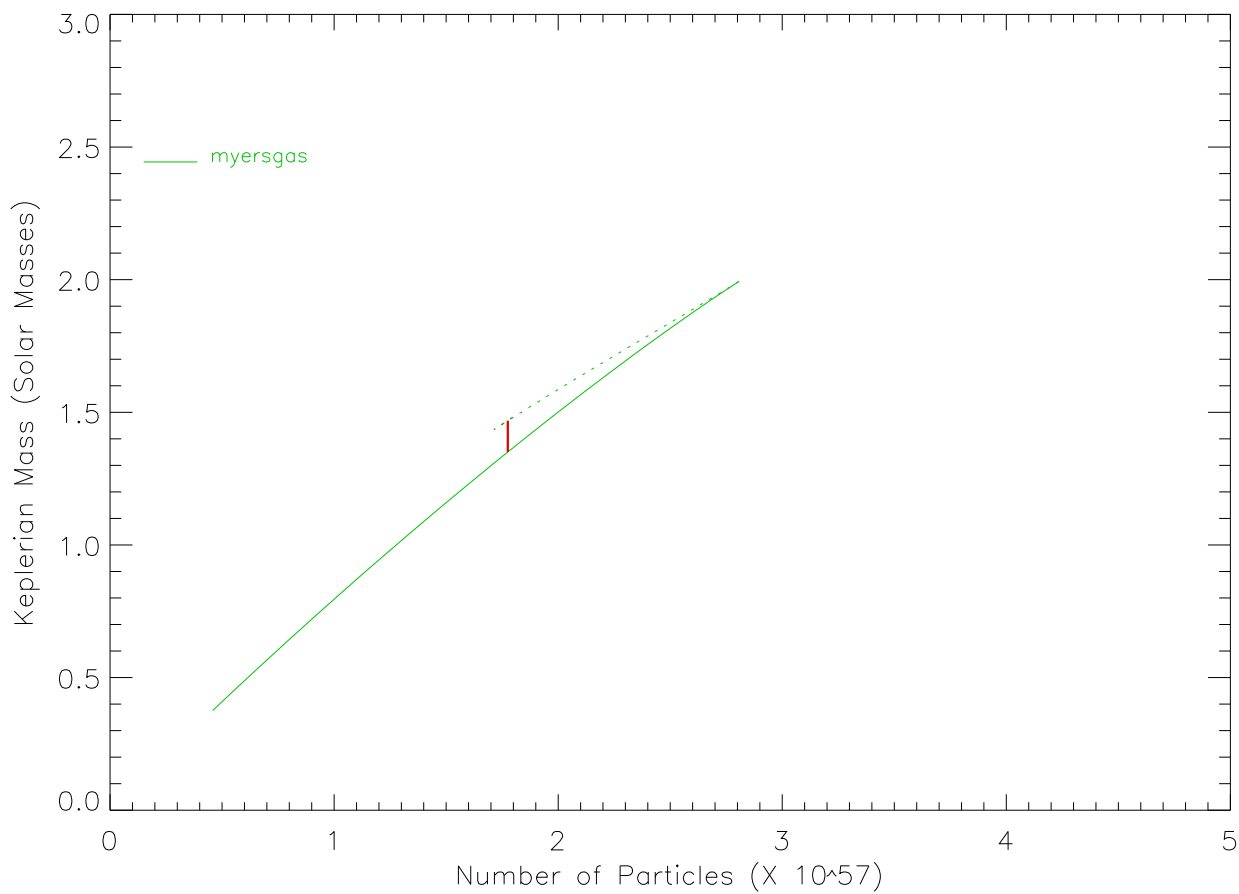}\caption{\emph{Keplerian Mass, M, vs. Total Baryon Number, A, of a self-gravitating
sphere in hydrostatic equilibrium, excerpt.} A single curve is extracted
from Figure \ref{fig:M-vs-A} in order to clarify discussion. The
vertical interval highlighted in red on this figure marks the difference
in mass between the stable and unstable equilibrium configurations
of an arbitrary number of baryons. The energy equivalent of the mass
difference is equal to the net work needed to compress the stable
configuration into the unstable configuration.\label{fig:M-vs-A-excerpt}}
\end{figure}

For clarity we extract one curve from Figure \ref{fig:M-vs-A} and
display it in Figure \ref{fig:M-vs-A-excerpt}. Note that central
density increases along the stable (solid line) portion of the curve
from lower left to upper right, and then increases still more along
the unstable (dotted line) portion of the curve. Due to the strongly
damped oscillations in both baryon number and Keplerian mass, extending
this curve (or any other curve from Figure \ref{fig:M-vs-A}) even
to infinite central density would result in only a minute zig-zag
pattern at the lower left of its unstable branch, terminating at a
finite mass and baryon number barely distinguishable from the endpoint
shown.

We can examine the compression process using the relationship between
different configurations with the same baryon number as mentioned
above. The vertical interval demarcated on Figure \ref{fig:M-vs-A-excerpt}
is bounded by the stable and first unstable hydrostatic equilibrium
configurations of the same baryon number for a given EOS. This vertical
distance is the mass equivalent of the net work (Pressure-volume work
minus gravitational potential energy released) required to compress
the stable configuration to the point of gravitational collapse. One
such relationship between an arbitrary stable configuration and the
corresponding compressed, unstable configuration with the same baryon
number is illustrated in Figure \ref{fig:M-vs-A-excerpt} with a bold
vertical line.

Consider the process of quasistatically compressing an ensemble of
matter from its stable hydrostatic equilibrium configuration to its
second hydrostatic equilibrium, this one unstable. The equilibrium
configurations at either end of the process must necessarily both
have pressures that vanish approaching their surfaces. Otherwise,
the finite outward pressure on the surface layer of the sphere would
cause it to expand into the vaccuum immediately (i.e., not in response
to a stochastic or microscopic perturbation), violating the definition
of hydrostatic equilibrium. The intermediate configurations (same
total baryon number and central density lying somewhere between the
central densities of the equilibrium configurations), however, cannot
possibly have pressures vanishing near their surfaces because these
configurations are in equilibrium with the external force compressing
the configuration. Figure \ref{fig:M-vs-A-excerpt} demonstrates that
an initially constrained, intermediate configuration, once allowed
to slowly relax and radiate mass-energy content, will come deterministically
to rest by expanding to the stable equilibrium configuration.

A simple analogy for the behaviors just described is that of a boulder
lying in a hilly region. A stable equilibrium configuration corresponds
to a boulder location at the bottom of a valley. Any small amount
of work done on it will displace the boulder from its original location.
A force is then required to maintain the boulder's displaced position,
and if it is removed, the boulder will relax back to its initial location.
A large amount of work done, however, can displace the boulder to
the top of the nearest hill, from which it may return to its original
position or roll down the other side of the hill, depending on the
details of its perturbation from the top.

The boulder analogy highlights several important observations regarding
self-gravitating spheres in hydrostatic equilibrium. The two equilibrium
positions, stable and unstable, are self-contained in the sense that
they require no containment pressure and therefore have a zero pressure
boundary, corresponding to zero force required to maintain the boulder's
position at the top and bottom of the hill. In the intermediate positions,
however, a containment pressure is required to maintain the structure
of the sphere, just as a force is required to hold the boulder in
place on the side of the hill. Because the containment pressure required
to maintain or enhance the compression of the sphere is known to be
zero at either equilibrium configuration, yet positive in between,
continuity requires that there be a maximum containment pressure somewhere
in the middle. Analogously, the hill must rise from a flat valley
to a maximum steepness, then return to a flat hilltop.

Like a boulder at the top of a hill, once a self-gravitating sphere
in hydrostatic equilibrium has been compressed from its stable equilibrium
position to its first unstable equilibrium, it may follow one of two
different paths depending on the details of its perturbation. The
sphere will either gravitationally collapse or expand back to its
stable configuration. All components of an evolving arbitrary perturbation
corresponding to oscillatory eigenmodes will soon be negligible compared
to the exponential growth eigenmodes. Moreover, unless the initial
perturbation is fine-tuned, the fastest growing mode, i.e. the mode
with the largest magnitude of frequency, is likely to soon dominate
all other modes. In the case of self-gravitating, hydrostatic spheres,
growth modes are determined by the sign of the square of their frequencies,
so the largest (imaginary) frequency will correspond to the lowest
(real) frequency squared- our fundamental mode, as discussed in Section
\ref{sub:Stability Criterion Theory}. Recall that it is a pure expansion
or contraction when that mode grows rather than oscillates, with no
nodes except at the center, as required by symmetry. Whether the self-gravitating
unstable hydrostatic sphere returns to its original, stable configuration
or collapses most likely depends, then, solely on whether the initial
phase of the {}``breathing'' mode corresponds to a contraction or
expansion.

\begin{figure}[H]
\includegraphics{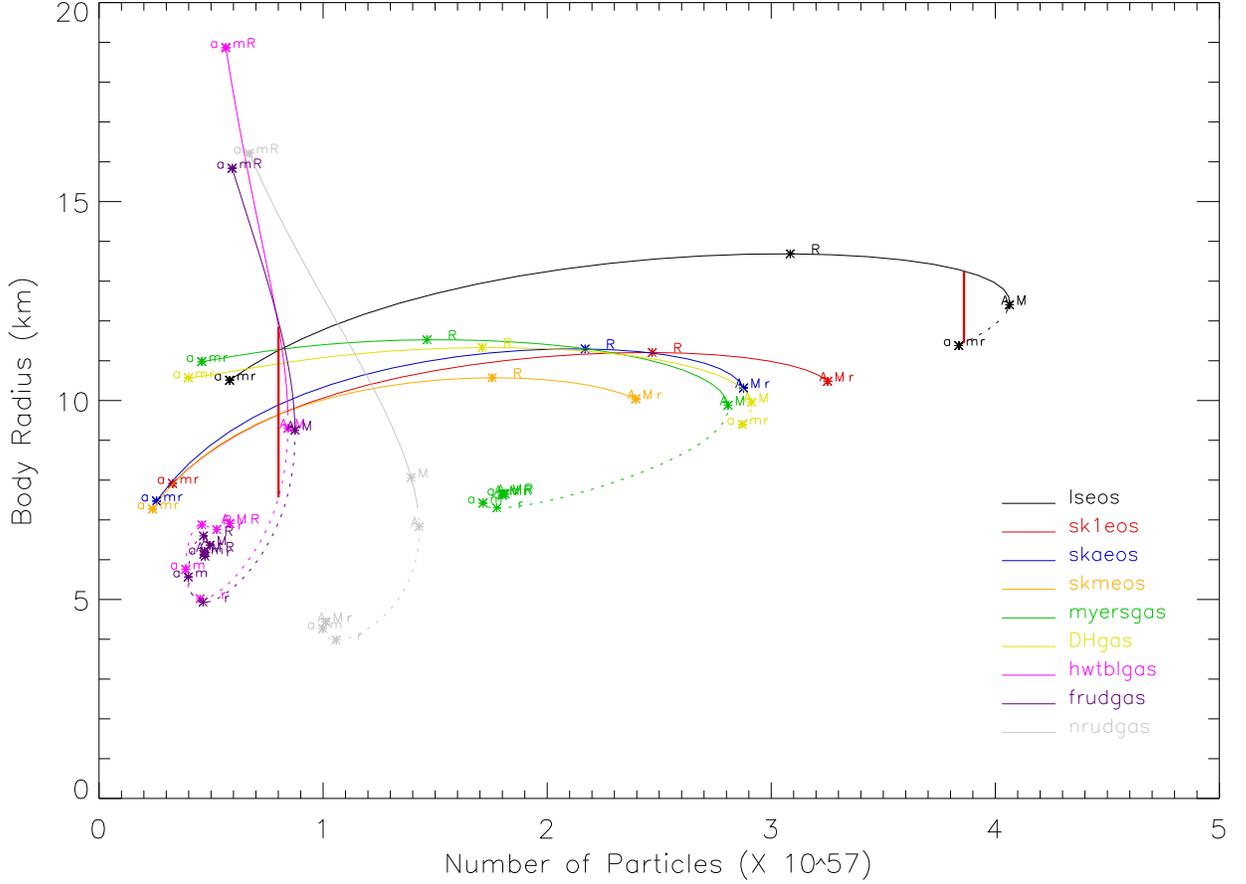}

\caption{\emph{Radius, R vs. Baryon Number, A, of a self-gravitating sphere
in hydrostatic equilibrium.} In order to perturb a self-gravitating
sphere in stable hydrostatic equilibrium, one must compress the sphere
from the larger radius indicated by a solid curve vertically downwards
to the corresponding unstable equilibrium point. Soft EOSs require
much compression to go unstable. This is illustrated by the long vertical
line connecting the stable configuration of 0.8 x 10$^{\text{57}}$
baryons in the HWTblGas EOS to the unstable configuration of the same
number of baryons in the same EOS. Stiff EOS require a much smaller
change in radius. This is illustrated by the short vertical line connecting
the stable and unstable configurations of 3.8 x 10$^{\text{57}}$
baryons in the LS EOS. Both of these baryon numbers correspond to
95\% of the baryon number of the configuration at the TOV limit.\label{fig:R-vs-A} }
\end{figure}

In Figure \ref{fig:R-vs-A}, with vertical lines we show the change
in radius necessary to bring two example configurations from stable
to unstable equilibrium. We see that the softer EOSs such as HWTblGas
require much greater compression than the stiffer EOSs, such as LS.
We have drawn our comparisons at 0.7 and 3.9 x 10$^{\text{57}}$ baryons,
respectively. These baryon numbers are both 95\% of the baryon number
of the equilibrium configuration at the TOV limit for that EOS.

\subsection{Generalization of the Theorem\label{sub:Stability Criterion Generalization}}

Chandrasekhar's stability theorem as stated in HTWW has two limitations
within which we must work: the theorem is only applicable to spheres,
and it is assumed that those spheres are surrounded by vaccuum. Because
gravity interferes only constructively, if a given sphere in hydrostatic
equilibrium is unstable to gravitational collapse, then any arbitrary
mass distribution containing a spherical region everywhere denser
than that given sphere will also be gravitationally unstable. We implement
this observation by considering only spherical regions excerpted from
within computer simulation data sets in order to detect zones of gravitational
collapse. We then estimate the mass contained within the spherical
region as the lower limit of mass of a DBH formed from that zone of
collapse. This only makes our criterion more conservative.

\begin{figure}[H]
\emph{\includegraphics{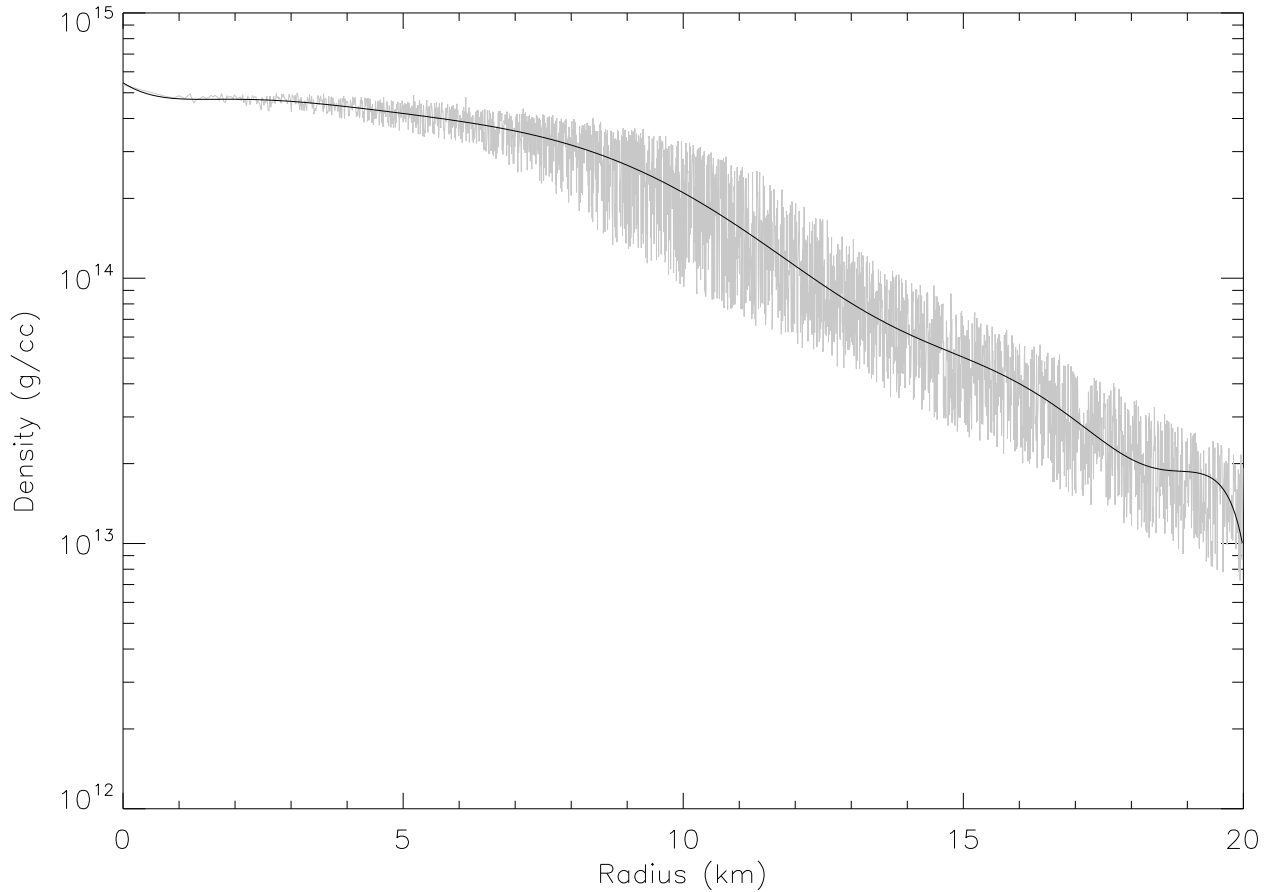}}\caption{\emph{Radial Profile of a Typical Spherical Region Being Analyzed
for Stability.} This profile comes from the Fryer supernova simulation
discussed in detail in Section \ref{sub:Type-II-Supernova}. For economy,
one out of every ten data points extracted from simulation data are
shown plotted in gray. The tenth-order polynomial fit to the full
ensemble of extracted data points is overplotted in black. This region
comprises roughly one half of a solar mass. The Type I technique developed
in this paper, with this EOS, can detect DBHs between 1.5 and 2 $M_{\odot}$,
so this region is ruled out as a Type I DBH progenitor. We discusss
the detection of Type II DBHs, which extend to arbitrarily small masses,
in Sections \ref{sub:Smaller-Still-DBHs} and \ref{sub:Detecting-Higher-Order-Instabilities}.\label{fig:Scoop Profile}}
\end{figure}

Because simulation data are very noisy, we are forced to take an average
over angles for a given inscribed spherical region in a simulation
data set. In Figure \ref{fig:Scoop Profile} we plot the density profile
thus generated from a region we examined for instability. This procedure
is discussed in more detail in Section \ref{sub:Algorithm}. This
approximation is not a strictly conservative assumption, so it may
introduce false positives. These false positives may be weeded out
by conducting a post-processing, ultra-high-resolution mini-simulation
focused on the region in question, using the original data set as
a spatial boundary condition throughout the mini-simulation.

In Figure \ref{fig:Scoop Profile} we show the region around the densest
data point in the Fryer Type II supernova simulation data set. To
generate this radial profile, all data points within 20 km are extracted
and sorted by distance from the densest data point. The profile shown
is roughly one half of a solar mass. Type I DBHs, with this EOS, are
between 1.5 and 2 $M_{\odot}$, so this region is not a Type I DBH
progenitor. Its central density also precludes it from being a Type
II DBH progenitor, but that diagnosis is sensitive to the simulation's
spatial resolution at the center of the region. We discuss an avenue
for detecting arbitrarily small, Type II DBHs in Sections \ref{sub:Smaller-Still-DBHs}
and \ref{sub:Detecting-Higher-Order-Instabilities}.

To address the second limitation of Chandrasekhar's stability theorem,
the vaccuum boundary condition, we compare a non-vaccuum bounded spherical
region extracted from a simulation data set to a vaccuum-bounded theoretical
model sphere known to be unstable via Chandrasekhar's theorem. Since
gravity only constructively interferes, if the spherical region of
data is everywhere denser than the unstable sphere, the spherical
region will also be unstable.

\subsection{Smaller-Still Dwarf Black Holes?\label{sub:Smaller-Still-DBHs}}

The work we have presented thus far deals with spherical configurations
only so compacted as to have a single mode of instability. For very
large baryon numbers, greater than the maxima shown in Figure \ref{fig:A-vs-rho0},
no stable equilibrium configuration exists- this is the familiar result
that, above the very restrictive TOV limit of at most a few solar
masses \citep{Heavy NS}, material can only be supported by an active
engine in the interior (e.g. fusion, for stellar bodies), and when
that engine shuts down, catastrophe ensues. Counterintuitively, there
is also a minimum baryon number for finding an unstable equilibrium
state in the manner previously described. For any given EOS, the first
minimum in baryon number or Keplerian mass vs. central density after
the maximum signifying the TOV limit is the absolute minimum for all
higher central densities. The curves shown in Figures \ref{fig:A-vs-rho0},
\ref{fig:M-vs-rho0}, and \ref{fig:R-vs-rho0}, if extrapolated arbitrarily
while maintaining constraints of physicality, would be characterized
by damped oscillation in baryon number, mass, and radius, respectively,
with increasing central density. For instance, from Figure \ref{fig:M-vs-rho0},
we can see for the Myers and Swiatecki EOS, no vaccuum-bounded, unstable
configuration of mass 1 $M_{\odot}$ exists. Yet, if one slowly compresses
a small amount of matter to an arbitrarily high pressure, a gravitational
collapse must ensue sooner or later. What, then, does $\sim$1 $M_{\odot}$-worth
of material look like, just before it is compressed to the point of
gravitational collapse?

Since we have posited that the extreme DBH progenitors result from
compressing a small amount of material, let us consider a small, isolated,
hydrodynamically stable, static, spherical configuration of matter
bounded by vaccuum. We then introduce outside that configuration (through
sufficiently advanced technology) a rigid, impermeable, spherical
membrane whose radius can be changed at will. If we gradually shrink
the membrane, even to the verge of the gravitational collapse of its
contents, then the configuration of matter inside must at all times
obey Equations \ref{eq:dmdr} and \ref{eq:dPdr}. This is because
the interior layers of the matter inside the membrane are insensitive
as to what is causing the layers above them to exert a particular
inward pressure.

The structure equations are first order in $r$, so if two solutions
are found with the same (central) boundary condition, they must be
identical from the center to the radius of the membrane. The progenitors
of extreme DBHs (when constructed quasi-statically), therefore, are
truncated solutions of the structure equations, with central density
high enough that the complete solution (out to sufficient radius for
pressure and density to reach zero) would have two or more radial
eigenmodes that grow rather than oscillate. In Table \ref{tab:Mass-Range-Scenarios}
we have compiled the scenarios under which different amounts of matter
would collapse, gravitationally.

\begin{longtable}{|>{\centering}m{1in}|>{\centering}m{1in}|>{\centering}m{4in}|}
\caption{\emph{Mass Ranges of Qualitatively Different Gravitational Collapse
Scenarios} Any amount of matter is vulnerable to gravitational collapse
if it becomes sufficiently dense. Different amounts of matter will
exhibit qualitatively different gravitational collapse scenarios.
The numerical figures of 3/2 and 2 M$_{\odot}$ are approximate and
derived from the \citet{MS} EOS and the \citet{Heavy NS} neutron
star. That neutron star is assumed to be very near the TOV limit since
it is so much more massive than any other yet observed. That EOS is
chosen because its TOV limit is the closest to 2 M$_{\odot}$, and
its first post-TOV-limit mass minimum gives rise to the 3/2 M$_{\odot}$
number.\label{tab:Mass-Range-Scenarios}}
\tabularnewline
\hline 
Mass Range & Exemplar & Discussion\tabularnewline
\hline
\endfirsthead
\hline 
Mass Range & Exemplar & Discussion\tabularnewline
\hline
\endhead
\hline 
> 2 M$_{\odot}$ & Direct collapse without supernova of a very massive star & This mass range lies above the TOV limit of the real-world, universal
EOS. There is no permanent equilibrium configuration for this amount
of matter whatsoever. Thus it can only be supported in short-term
hydrostatic equilibrium by an active (i.e., explicitly \emph{out}
of chemical- or other-equilibrium) engine consuming some sort of fuel
on a longer characteristic time scale, such as fusion in massive stars
or energy released by a black hole capturing matter, as in AGNs. When
the fuel is expended, catastrophe will inevitably ensue.\tabularnewline
\hline 
2 M$_{\odot}$ & Collapse of accreting neutron star & This is the collapse of an object that has reached precisely the TOV
limit by growing quasi-statically, as through slow accretion. Catastrophe
ensues with the accretion of one more particle. The approximate value
of this mass is taken from \citet{Heavy NS}, assumed to be very near
the TOV limit.\tabularnewline
\hline 
$\frac{3}{2}$ - 2 M$_{\odot}$ & Type I DBH formed in, perhaps, hypernova & A configuration of matter in this range will assume a stable equilibrium
configuration. If it is then slowly compressed by exterior force,
it will eventually reach a second equilibrium, this time unstable.
As the configuration approaches that second equilibrium, the external,
boundary pressure decreases, and vanishes entirely upon reaching it.
It will collapse if perturbed in a manner dominated by compression,
or expand to its stable equilibrium configuration if perturbed in
a manner dominated by expansion. The limits of this mass range are
defined by the region in which the curve of an assumed EOS decreases
in Figure \ref{fig:M-vs-rho0}, to the right of the TOV limit. We
give a value of $\frac{3}{2}$ M$_{\odot}$ because that is the approximate
limit corresponding to the myersgas EOS, the EOS we examined with
a TOV limit closest to \citet{Heavy NS}.\tabularnewline
\hline 
< $\frac{3}{2}$ M$_{\odot}$ & Type II DBH formed in, perhaps, SN Ia & A configuration of matter in this range will settle into a stable
equilibrium configuration. It will exhibit no second, unstable equilibrium
configuration, for any amount of compression it is forced to undergo.
Instead, if compressed to the point of collapse, its core will commence
doing so even while its upper layers continue to exert a pressure
against the container compressing it.\tabularnewline
\hline
\end{longtable}

To physically realize a truncated mathematical solution, we have no
choice but to constrain a configuration with something functionally
equivalent to the membrane of the above paragraphs. In the supernova
simulations in which we would like to find extreme DBHs, this role
would be fulfilled by turbulence or fallback driving mass collisions.
In other words, the potentially unstable region is surrounded by high
pressure gas acting to compress it.

Because they are truncated before the radius where pressure reaches
zero, the progenitors of extreme DBHs will not appear in the figures
depicting self-gravitating spheres in hydrostatic equilibrium, which
have containment pressures of zero. As described in Section \ref{sub:The-Fate-of-unstable},
the extreme DBH progenitor spheres would expand if the membrane (or,
realistically, the surrounding gas) were withdrawn, but under the
influence of the membrane they can achieve instability to gravitational
collapse.

HTWW (p. 46) demonstrate that the zero-containment-pressure configurations
lying in the first unstable region with increasing baryon number of
Figure \ref{fig:A-vs-rho0} have two growth eigenmodes. The second
eigenmode will have a single node, and in the case of two growth eigenmodes,
this enables the outer layers to expand while the inner layers contract
catastrophically. This is the collapse scenario by which very small
DBHs could form. Unfortunately, detecting this instability is a far
less tractable problem than the work we have already presented, instabilities
arising in zero-containment pressure configurations from the growth
of the lowest eigenmode. It is less tractable because in order to
reach a definite conclusion about stability we must be able to calculate
the radius necessary to bring a Type II DBH progenitor to the brink
of collapse, or equivalently, calculate where we must truncate a mathematical
solution to the structure equations (\ref{eq:dmdr}-\ref{eq:dPhidr})
in order to leave a configuration only marginally stable. In contrast,
for Type I DBHs, the critical radius arises from the integration of
the structure equations automatically.

The complete, equilibrium sphere solutions from which we obtained
our truncated solutions lie to the right of the first unstable minimum
in Figure \ref{fig:A-vs-rho0}. This location means that they have
two or more eigenmodes of exponential growth. Our thought experiment
of slowly compressing small amounts of matter, however, demonstrated
that the truncated solutions, on the other hand, can have only a single
mode of growth. That is the very criteria by which we decided to halt
our thought-compression; we stopped as soon as the sphere became unstable
to gravitational collapse, and that requires only a single growth
mode (Section \ref{sub:Stability Criterion Theory}). We find, therefore,
that truncating a solution apparently changes its eigenspectrum.

This may seem strange. Why shouldn't the truncated, inner layers simply
collapse in the same manner as the whole sphere would? How do the
inner layers know and respond to the behavior (or even existence)
of the upper layers? Considering the boundary condition resolves the
puzzle. The untruncated solution can expand into the vaccuum in response
to an outgoing perturbation, so the outer boundary condition is transmissive.
The kinetic energy of the perturbation would be gradually converted
to potential energy or radiated entirely. It could return to the center
only as the entire sphere relaxed. The truncated solution, in contrast,
is abutted by a physical obstacle that may reflect some perturbation.
Some of the kinetic energy of a perturbation of such a sphere would
rebound directly to the center. Just as in the simple classroom demonstration
of changing the pitch of a pipe organ pipe by opening and closing
its end, changing the boundary condition of a self-gravitating sphere
changes its eigenspectrum. To detect these instabilities, we must
decide where to truncate a solution so that it includes just enough
material to be unstable. Our calculation to determine where to truncate
follows in the next section.

\subsection{Detecting Higher-Order Instabilities\label{sub:Detecting-Higher-Order-Instabilities}}

\begin{figure}[H]
\includegraphics{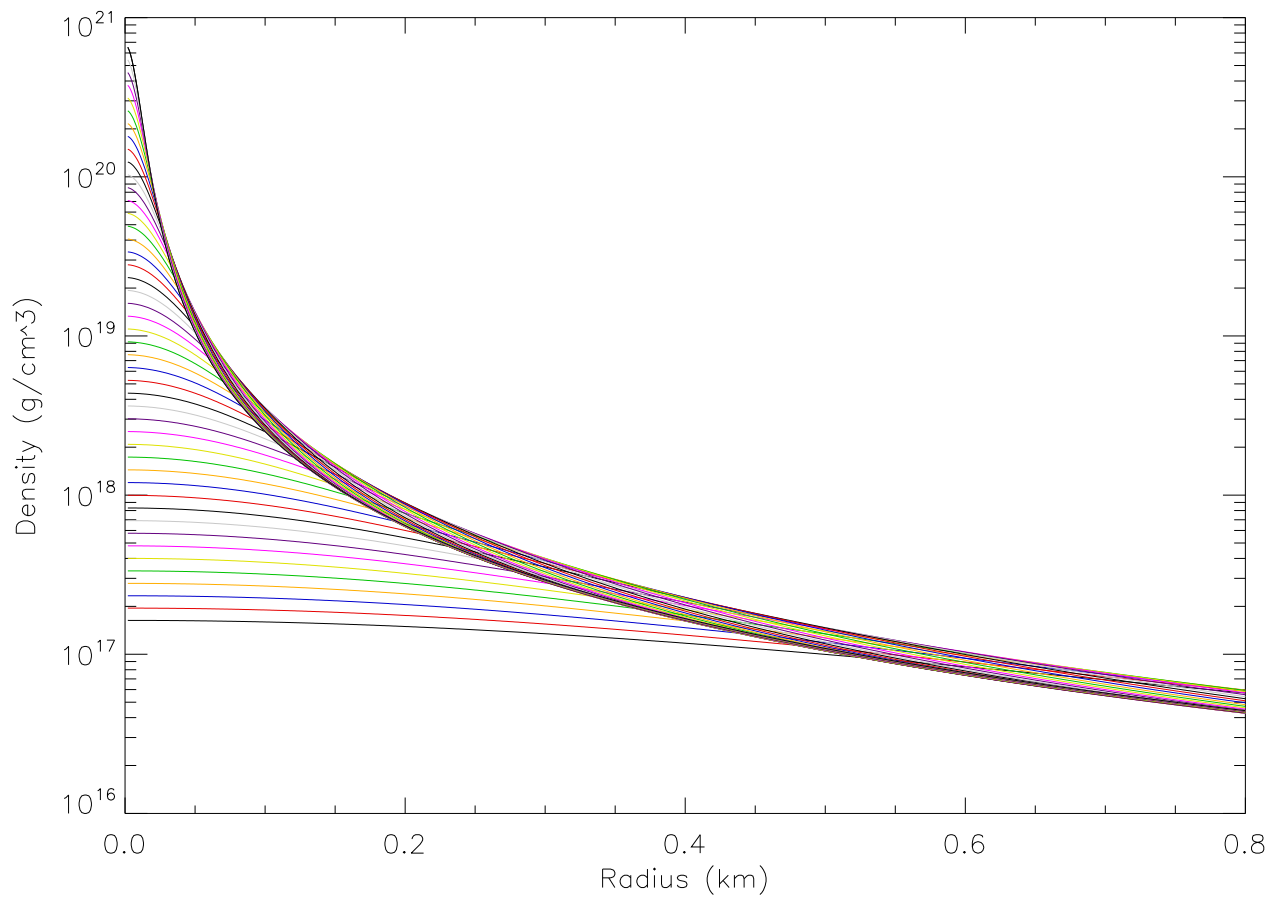}\caption{\emph{Central radial density profiles of extreme self-gravitating
spheres in hydrostatic equilibrium.} These curves, plotting the innermost
regions of Type II DBH progenitors, show that these progenitors may
not be detectable in simulation data at all unless the resolution
is extremely fine, perhaps smaller than 100 m, because even the most
and least extreme Type II DBH progenitors are nearly indistinguishable
outside 600 m. The curve with lowest central density is the first
Type II DBH progenitor, i.e., the first curve that, taken all the
way to $P=0$, displays two or more modes of instability. This figure
extrapolates the myersgas EOS much farther than it is intended to
be taken, but any physically realistic EOS will display the same qualitative
behavior.\label{fig:Peakiness}}
\end{figure}

The marginally stable progenitors of Type II DBHs have very high central
densities. From Figure \ref{fig:A-vs-rho0} we see that for the Myers
and Swiatecki EOS, for instance, the onset of instability of a second
mode (first minimum after first maximum) occurs at a central density
roughly 30 times greater than the onset of instability of the first
eigenmode (first maximum). For this EOS, these critical points are
at central densities of $~$$\sim$10$^{\text{17}}$ g cm$^{\text{-3}}$
and $\sim$$~$3 $\times$ 10$^{15}$ g cm$^{\text{-3}}$, respectively.
As will be discussed in more detail in Section \ref{sub:Type-II-Supernova},
the highest density we have yet found in a simulation is $~$$\sim$7
$\times$ 10$^{\text{14}}$ g cm$^{\text{-3}}$. We are currently
seeking data sets achieving higher densities. Happily, as shown in
Figure \ref{fig:Peakiness}, the radial profiles of Type II DBH progenitors
are very sharply peaked. The required region of extreme, thus-far-unprecendented
density may be very small indeed. The curve shown in this figure with
the lowest central density is the first Type II DBH progenitor, i.e.,
the first hydrostatic sphere that has more than one growth eigenmode.

\begin{figure}[H]
\includegraphics{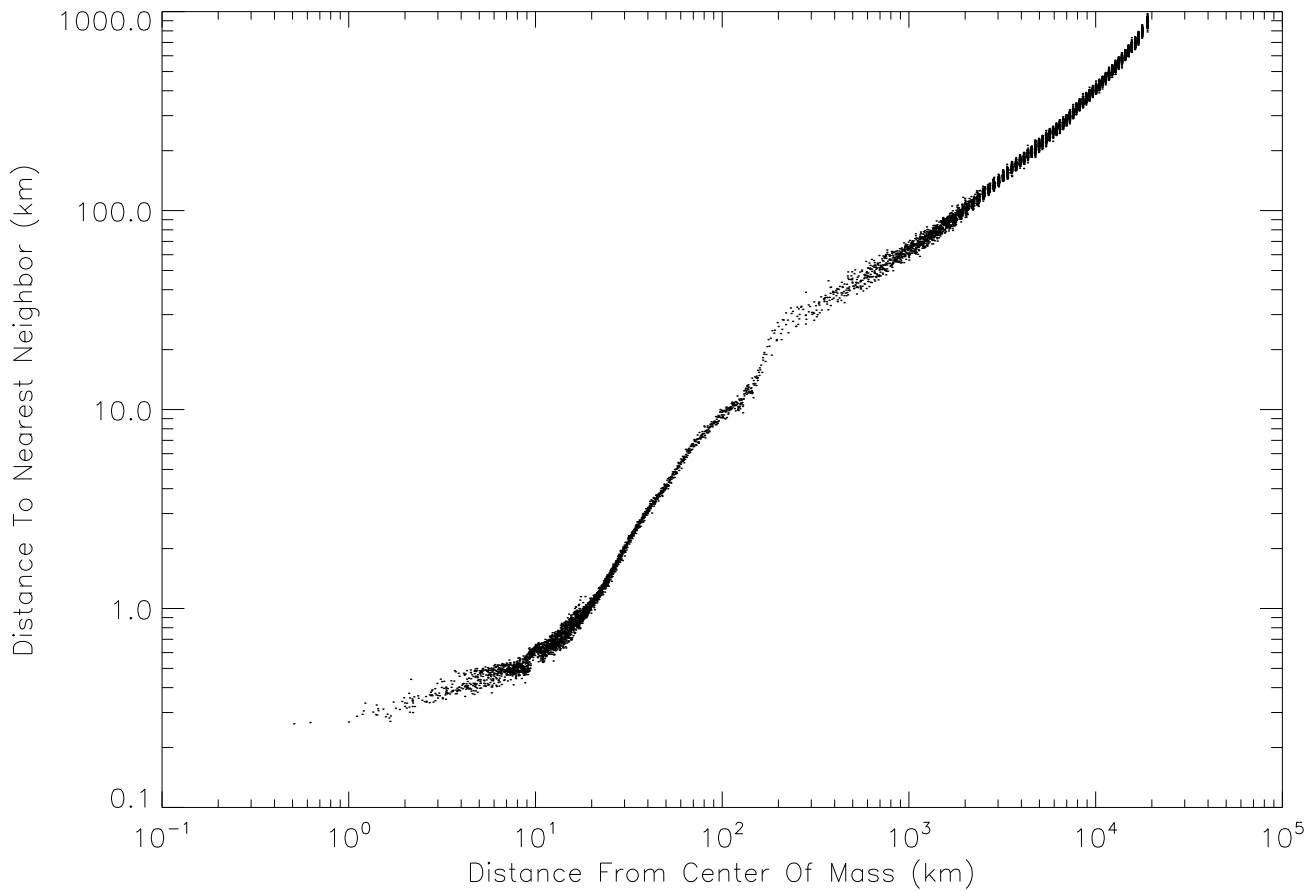}

\caption{\emph{Distance to Nearest Neighbor vs. Distance from Center of Mass
for the Fryer Type II Supernova Simulation} For economy we plot the
distance to nearest neighbor for 1/10 of the data points in the Fryer
data set. Figure \ref{fig:Peakiness} demonstrated the need for simulations
of very high resolution, of order 100 m. Even with the computing resources
available at Los Alamos National Laboratory that resolution is only
found within $\sim$10 km of the center of mass of this simulation.\label{fig:Distance-to-Nearest-Neighbor} }
\end{figure}

In Figure \ref{fig:Peakiness} we see that the curves plotted that
have the highest central density decrease in density by three orders
of magnitude within roughly 200 m, and even the most and least extreme
curves plotted on this chart are virtually indistinguishable outside
600 m. DBH progenitors have no chance at all of being resolved in
a simulation data set except in regions where the spatial resolution
is on the order of 100 m or less. In Figure \ref{fig:Distance-to-Nearest-Neighbor}
we plot the distance to the nearest neighboring data point for a sampling
of data points in the Fryer Type II Supernova we analyzed. This simulation
was conducted recently using the full computing resources of Los Alamos
National Laboratory, yet even it achieves the necessary resolution
only for a 10 km radius around its center of mass, or less.

The high-central density termination point of the lseos, sk1eos, skaeos,
skmeos, and DHgas curves in Figures \ref{fig:A-vs-rho0}, \ref{fig:M-vs-rho0},
\ref{fig:M-vs-A}, \ref{fig:M-vs-A-excerpt}, and \ref{fig:R-vs-A},
are defined by the upper limits of published tables. In several of
those cases we see very little of even the single growth-eigenmode
regime, much less the two growth-eigenmode. (The myersgas, hwtbl,
frudgas, and nrudgas EOSs are published or known in closed form and
thus their curves are extended indefinitely, but in configurations
with central densities much beyond roughly 8 $\times$ 10$^{\text{14}}$
g cm$^{\text{-3}}$, that is, twice nuclear density, these should
be regarded as ambitious extrapolations suitable for illustrative
purposes only.) For confidence in calculating the properties of DBH
quasi-static progenitors, the upper limits of present EOSs must be
extended.

Let us optimistically suppose some time in the future we will have
an EOS reliably applicable to ambitious extremes, and that high-resolution,
future supercomputer simulations of extreme phenomena reveal very
high densities. How then should we proceed, i.e., could we generate
some theoretical compressed and constrained configurations on the
verge of higher-order acoustical collapse to compare to our hypothetical,
hydrodynamic supercomputer simulation? This suggested procedure is
in analogy to the algorithm discussed in Section \ref{sub:Algorithm},
where theoretical curves describing Type I DBH progenitors are compared
to spherical regions of existing simulation data.

We must conduct a thought experiment in order to discover more detail
about the progenitors of Type II DBHs. For concreteness, let us use
the myersgas EOS and assume it is absolutely accurate to all densities
with infinite precision. Let us consider two ensembles of matter of
slightly different total masses. Let both lie below the TOV limit,
roughly 2 M$_{\odot}$, so a stable equilibrium configuration exists
for both. Allow the more massive ensemble to come to hydrostatic equilibrium,
and spherically compress the less massive ensemble through outside
forces so that its central density is equal to the central density
of the first ensemble. Because the structure differential equations
governing hydrostatic equilibrium (Equations \ref{eq:dmdr} and \ref{eq:dPdr},
along with their adjuncts, Equations \ref{eq:dAdr} and \ref{eq:dPhidr})
are first order (in radius) and we are maintaining equal (central)
boundary conditions by fiat, both ensembles will have identical radial
dependence of their physical quantities. The only difference between
the two is that the more massive ensemble extends from $r=0$ to $R$,
that is, the radius where pressure vanishes, while the less massive
ensemble truncates at $r\lesssim R$, where pressure is still finite.

Let us then quasistatically compress both of our spherical ensembles,
maintaining equal central density between the two. That central density
will increase monotonically. The more massive ensemble will reach
the verge of gravitational collapse first, while the less massive
ensemble will need to be compressed to a slightly higher density to
gravitationally collapse. Since we have made the masses of the two
ensembles arbitrary apart from being below the TOV limit, we have
proven the lemma that:

\emph{1. The central density of a static sphere on the verge of gravitational
collapse increases monotonically as the total mass of the sphere decreases.}

This statement applies to progenitors of both Type I and Type II DBHs.
We already know the central density objects of roughly 2 M$_{\odot}$
down to 1.5 M$_{\odot}$ (in the myersgas EOS) on the verge of collapse
(i.e., Type I DBH progenitors, which lie between the first maximum
and following minimum on Figure \ref{fig:M-vs-rho0} of a curve corresponding
to a selected EOS) will have central densities of roughly 2 $\times$
10$^{\text{15}}$ erg cm$^{\text{-3}}$ - 10$^{\text{17}}$ erg cm$^{\text{-3}}$,
respectively. To look for objects of smaller mass on the verge of
collapse, we must increase the central density even more.

\emph{2. A Type II DBH progenitor will have a central density greater
than the minimum following the first maximum of an EOS's curve as
shown in Figure \ref{fig:M-vs-rho0}.}

Type I DBH progenitors exert a vanishing pressure on their containers
at the end of their compression because they are then identical to
full, vaccuum-bounded solutions to the structure equations. If we
look beyond central densities of 10$^{\text{17}}$ erg cm$^{\text{-3}}$
on Figure \ref{fig:M-vs-rho0}, however, the vaccuum-bounded solutions
of the structure equations then commence increasing, while the mass
of Type II DBH progenitors decrease, by construction. Nevertheless,
since we arrived at these configurations by quasi-static compression,
the Type II DBH progenitors must obey the structure equations. To
reconcile this paradox, we have only one possible conclusion.

\emph{3. A Type II DBH progenitor is not described by the full, vaccuum-bounded
solution of the structure equations with corresponding central density.
It is a radial truncation of that solution at a smaller radius, where
the pressure is still finite and the mass enclosed is less than the
full solution.}

Let us call the radius at which an ensemble reaches the verge of collapse
the \emph{critical radius} for that mass and baryon number. HTWW (p.
46) proved that the full, vaccuum-bounded solutions of the structure
equations from which DBH progenitors are truncated have two or more
growth eigenmodes. From our construction in the thought experiment,
we dubbed a compressed sphere a DBH progenitor at the moment it acquired
its first growth eigenmode. We could therefore discover the value
of the critical radius for a particular central density by selecting
different test radii at which to truncate that solution. We would
then use a version of Chandrasekhar's stability theorem (Equation
\ref{eq:omega2}) generalized to non-zero exterior boundary pressure
to determine how many growth eigenmodes that truncation exhibits.
The critical radius lies at the boundary between zero growth eigenmodes
and one. Finally, repeating that procedure with structure equation
solutions of many different central pressures would allow us to derive
the relationships between other physical quantities of spheres on
the verge of collapse such as total mass, exterior pressure, and density
at the center and at the exterior boundary.

Unfortunately for us, Chandrasekhar's derivation \citep{Chandra},
being a calculus of variations technique, necessarily assumes the
existence of a node at the outer boundary of the spherical region
to which it is applied and therefore inextricably assumes vanishing
pressure at the outer radius of the body in question. However, since
we have not yet discovered any regions within simulations dense enough
to be comparable to the central density of Type II DBHs, we have concluded
that deriving a generalization of Chandrasekhar's theorem utilizing
non-zero exterior pressure boundary conditions is unnecessary at this
time.

\subsection{The Algorithm for Finding Dwarf Black Holes\label{sub:Algorithm}}

We summarize the results of our arguments for determining if there
exists a region of matter in a spacelike slice of a general relativistic
simulation that will collapse into a DBH as follows:
\begin{enumerate}
\item Find the densest data point inside a computational simulation of an
extreme phenomenon (e.g. supernovae as described above) and take a
spherical {}``scoop'' out of the data centered on the densest point.
For guidance on the size of scoops to remove, we can consult Figure
\ref{fig:R-vs-rho0}. For a given EOS, the size of the scoops we need
examine are limited to the range of radii from the radius of spheres
at the onset on instability, i.e. the junction between solid and dashed
curves, and the radius of spheres with central density equal to the
highest single data point in the simulation data set. %
\begin{figure}[H]
\includegraphics{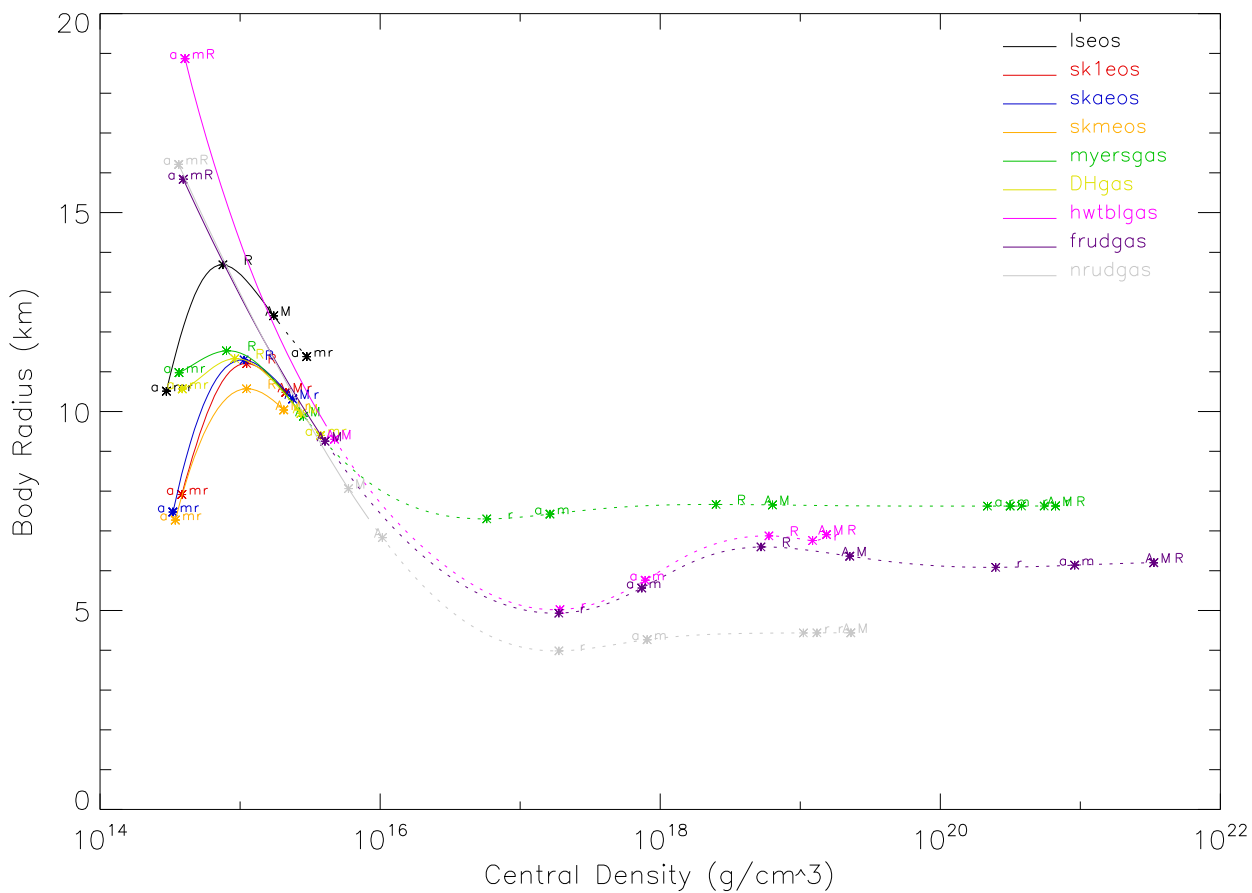}

\caption{\emph{Radius, R, vs Central Density, $\rho{}_{\text{0}}$, of a self-gravitating
sphere in hydrostatic equilibrium}. These curves are drawn with the
same conventions as those in Figure \ref{fig:A-vs-rho0}. We can observe
quantitatively the effects of having a {}``stiff'' equation of state.
Those EOSs that lead to a large estimate for the TOV limit also lead
to an increase in radius with increasing $\rho{}_{\text{0}}$ before
the final decrease in radius approaching the onset of instability.
{}``Soft'' equations of state lead to monotonic decreases in radius
as $\rho{}_{\text{0}}$ increases from the lower limit of this figure
to the onset of instability. The high central density limits of the
lseos, sk1eos, skaeos, skmeos, and DHgas are defined by the upper
limits of their published tables. The other curves are known in closed
form and are thus extended indefinitely, but beyond roughly twice
nuclear density these data points should be considered for illustrative
purposes only.\label{fig:R-vs-rho0}}
\end{figure}

\item Construct a sphere in hydrostatic equilibrium according to the structure
equations tailored to match the physical properties of the scoop as
closely as possible. The central pressure of the theoretical sphere
is set equal to the central pressure of the data scoop. The temperature
and other {}``$X$'' thermodynamic variables are, as functions of
$r$, fitted to the radial profile of the data scoop. Our experience
with supernova simulation data sets reveals that this radial profile
is quite noisy, so we make a tenth-order polynomial least-squares
fit for each $X$ parameter for each scoop using IDL's \texttt{poly\_fit}
routine.
\item Calculate the stability of the tailored hydrostatic sphere using Chandrasekhar's
theorem as described in Sections \ref{sub:The-Fate-of-unstable} and
\ref{sub:Stability Criterion Generalization} for Type I DBHs and
in Sections \ref{sub:Smaller-Still-DBHs} and \ref{sub:Detecting-Higher-Order-Instabilities}
for Type II DBHs. Proceed only if it is unstable.
\item Compare the tailored, unstable hydrostatic sphere to the scoop. If
the scoop is everywhere denser than the sphere, it follows that the
scoop should also be unstable to gravitational collapse. If the scoop
is denser than the sphere only for an interior fraction of the sphere's
radius, then the scoop may still be unstable, as per our discussion
in Sections \ref{sub:Smaller-Still-DBHs} and \ref{sub:Detecting-Higher-Order-Instabilities}.
For this to occur, the sphere must be dense enough to exhibit at least
two growth eigenmodes, i.e., two negative eigenvalues of oscillation
frequency squared. Conveniently, the algorithm we use for calculating
eigenvalues returns all of them simultaneously, so we have the second
eigenvalue ready at hand. We have not yet found any simulations where
this scenario comes into play.
\item Determine the total mass of the scoop inside the radius of the tailored,
unstable hydrostatic sphere, and include it in the mass spectrum of
black holes produced in this simulation.
\item Repeat these steps, through scoops of descending central density,
until the pressure at the center of a scoop is less than the lowest
central density necessary to achieve instability. That density is
the central density of the sphere of mass equal to exactly the TOV-limit.
This is a purely conservative assumption. Larger spheres with lower
maximum densities also can achieve instability to gravitational collapse,
but are not equilibrated, so we have no mathematical tools to prove
that they are unstable, other than a direct comparison to the Schwarzschild
radius for that amount of matter or the execution of a detailed hydrodynamic
simulation.
\end{enumerate}

\section{Applications}

\subsection{Discussion\label{sub:Application Discussion}}

We are acquiring and testing simulation data of various extreme astrophysical
phenomena from different research groups. For best results, it is
desirable to employ the same EOS when constructing a tailored, unstable
sphere as was used in constructing the {}``scoop'' of simulation
data (as described in Section \ref{sub:Algorithm}) to which it will
be compared. Due to the constantly evolving and patchwork nature of
EOSes implemented in real-world simulations, this is not always possible,
however. We have already completed the analysis of one simulation
data set, a three dimensional, Type II supernova simulation provided
to us by Chris Fryer of Los Alamos National Laboratory.

\subsection{Type II Supernova\label{sub:Type-II-Supernova}}

\begin{figure}[H]
\includegraphics{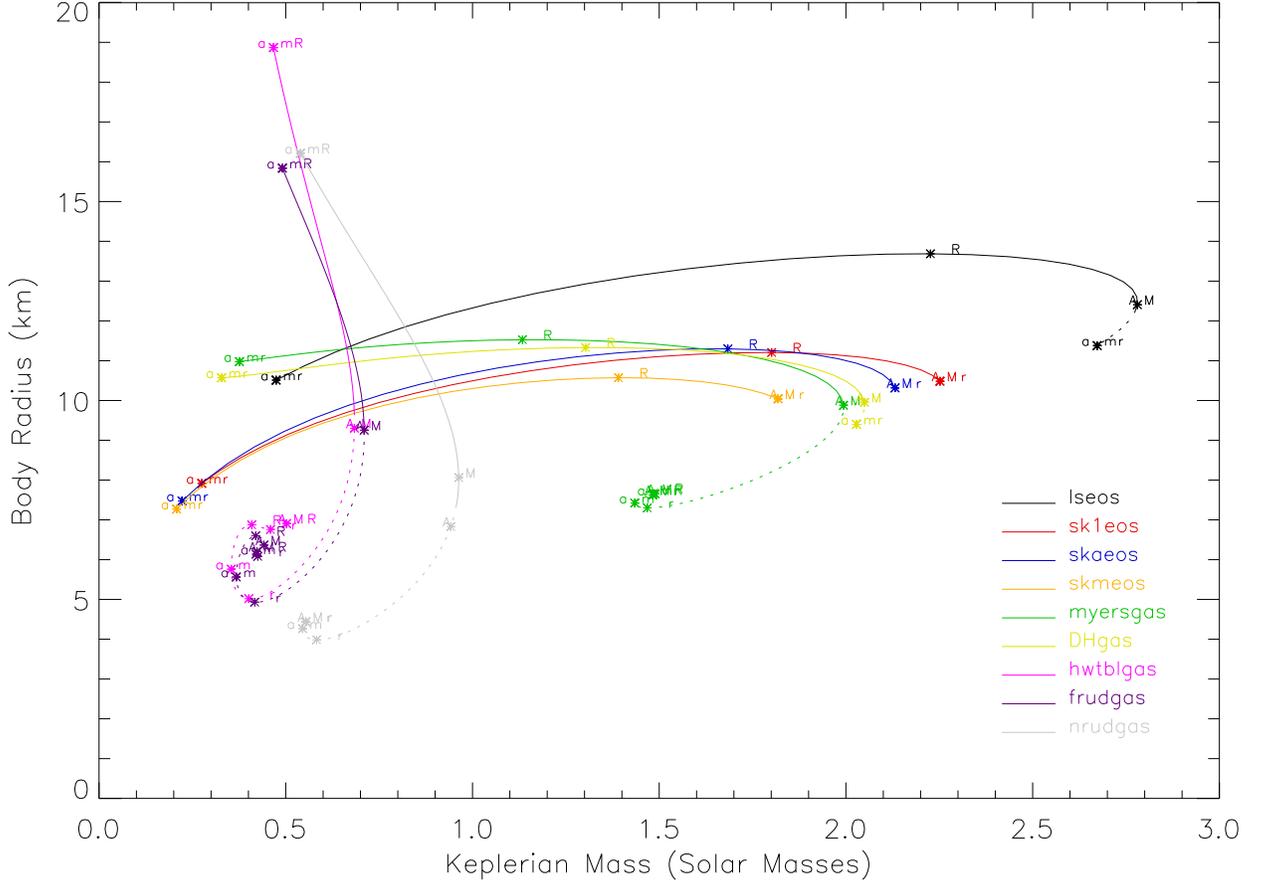}

\caption{\emph{Radius, R, vs. Keplerian Mass, M, of a self-gravitating sphere
in hydrostatic equilibrium.} This graph gives a minimum estimate of
the spatial resolution necessary for accurately detecting DBHs of
a given mass in a simulation data set. \label{fig:R-vs-M}}
\end{figure}

In order to determine whether a simulation has sufficient spatial
resolution to detect DBHs, we consult Figure \ref{fig:R-vs-M}, which
shows the radii of unstable spheres in hydrostatic equilibrium as
a function of Keplerian Mass. The $R$ vs. $M$ relation is also critical
to observational investigations of extreme phenomena, especially neutron
stars, as probes of the extreme EOSs. The simulation conducted by
Chris Fryer of the Los Alamos National Laboratory is fully three-dimensional,
and it easily has fine enough spatial resolution, especially in the
densest regions near the surface of the central proto-neutron star.

Fryer's simulation uses the SK1 EOS from the Lattimer-Swesty quartet
described in Section \ref{sub:Lattimer-&-Swesty-EOS}. This simulation
was evolved from the canonical 23 M$_{\odot}$ progenitor developed
by \citet{YA 2005 progenitor} using a discrete ordinates, smooth
particle hydrodynamics code. The progenitor code, TYCHO, is a {}``One-dimensional
stellar evolution and hydrodynamics code'' that {}``uses an adaptable
set of reaction networks, which are constructed automatically from
rate tables given a list of desired nuclei.'' \citep{YA 2005 progenitor}
The TYCHO code is {}``evolving away from the classic technique (mixing-length
theory) of modeling convection to a more realistic algorithm based
on multidimensional studies of convection in the progenitor star.''
\citep{FY2007} Fryer's supernova simulation is described in more
detail in \citet{FY2007}, particularly Section 2 of that paper.

The maximum density found in this simulation was 6.95 $\times$ 10$^{\text{14}}$
g cm$^{\text{-3}}$, while the minimum density necessary to create
a sphere in unstable equilibrium in this EOS is 2-3 $\times$ 10$^{\text{15}}$
g cm$^{\text{-3}}$. We do not have firm grounds on which to predict
the creation of dwarf black holes from Type II supernovae, but being
within a factor of 3-4 of the necessary density in one of the milder
members of the menagerie of extreme astrophysical phenomena suggested
we publish these results and proceed with investigating other supernova
simulations and simulations of other phenomena.

\subsection{Other Applications}

We are currently seeking data sets from other simulations to analyze
for possible DBH formation. One type of simulation in which we are
interested is those with large total mass, such as the death of very
massive stars or phenomena larger than stellar scale. We feel that
those types of simulation supply the best arena for the Type I (i.e.,
the most massive, least dense) DBHs to arise, as even a single specimen
of that end of the DBH spectrum would comprise a significant fraction
(or all!) of the ejecta from small-scale events. We are also interested
in smaller scale but extremely violent simulations, such as the explosive
collapse of a white dwarf or neutron star, or compact object mergers,
as these may give rise to the extremely high densities necessary for
Type II DBHs to form, even though they probably do not have enough
total mass involved in the event to create Type I DBHs.

Hypernovae, for instance, are a promising candidate for observing
Type I DBHs. In addition to the especially large amount of energy
released, hypernovae also occur in the context of very large ensembles
of matter, the better for carving off into somewhat subsolar mass
chunks. A recent study \citep{FYH HNova Prog} examined the collapse
and subsequent explosion of 23 $M{}_{\odot}$ and 40 $M{}_{\odot}$
progenitors. While this physical phenomenon may be a promising candidate
in the search for DBHs, the particular assumptions used by this group
to render the calculation tractable, while fully valid for their intended
research purposes, make it unlikely that we would find any DBHs in
their simulation of it.

First, their study was one-dimensional, thus precluding a detailed
treatment of aspherical phenomena including localized clumping into
DBHs. They also treat the densest regions of the simulation as a black
box with an artificially prescribed behavior, furthermore assumed
to be neutrino-neutral. Details critical to the formation of DBHs
may have been lost here. From \citet{FYH HNova Prog}:

\texttt{We follow the evolution from collapse of the entire star through
the bounce and stall of the shock... At the end of this phase, the
proto\textendash{}neutron stars of the 23 and 40 M$_{\odot}$ stars
have reached masses of 1.37 and 1.85 M$_{\odot}$, respectively.}

\texttt{At this point in the simulation, we remove the neutron star
core and drive an explosion by artificially heating the 15 zones ($\sim$3
\texttimes{} 10$^{\text{-2}}$ M$_{\odot}$) just above the proto\textendash{}neutron
star. The amount and duration of the heating is altered to achieve
the desired explosion energy, where explosion energy is defined as
the total final kinetic energy of the ejecta (see Table 1 for details). The
neutron star is modeled as a hard surface with no neutrino emission. Our
artificial heating is set to mimic the effects of neutrino heating,
but it does not alter the electron fraction as neutrinos would. The
exact electron fraction in the ejecta is difficult to determine without
multidimensional simulations with accurate neutrino transport. Given
this persistent uncertainty, we instead assume the electron fraction
of the collapsing material is set to its initial value (near Ye =
0.5 for most of the ejecta).}

In addition, the very material most critical in generating DBHs plays
no role in their investigation. Again, from \citet{FYH HNova Prog}:

\texttt{As this material falls back onto the neutron star its density
increases. When the density of a fallback zone rises above 5 \texttimes{}
10$^{\text{8}}$ g cm$^{\text{-3}}$, we assume that neutrino cooling
will quickly cause this material to accrete onto the neutron star,
and we remove the zone from the calculation (adding its mass to the
neutron star).}

They conclude their discussion with a very nearly prescient warning
to a reader interested in DBHs:

\texttt{Be aware that even if no gamma-ray burst is produced, much
of this material need not accrete onto the neutron star. Especially
if this material has angular momentum, it is likely that some of it
will be ejected in a second explosion. (In a case with no GRB this
will presumably be due to a similar mechanism, but without the very
high Lorentz factors.) This ejecta is a site of heavy element production
(Fryer et al. 2006). As we focus here only on the explosive nucleosynthesis,
we do not discuss this ejecta further.}

We are interested in the results of an updated study focused more
on the dynamics and/or ejecta of the supernova explosion and less
on nucleosynthetic yields, or one that preserved more of the information
that was irrelevant to Fryer, Young, and Hungerford's investigation.

\section{Conclusions}

We have proposed that a variety of extreme astrophysical phenomena,
some newly recognized to be more turbulent and aspherical than previously
appreciated, may compress local density concentrations beyond their
points of gravitational instability. Upon gravitational collapse and
expulsion, these local concentrations would become dwarf black holes
(DBH). We expect that, due to the enormous velocities with which e.g.
supernova ejecta is expelled, many or most DBHs may not be bound in
galaxies but will instead be found in the intergalactic medium. As
such, DBHs would not be subject to the constraints placed upon the
abundance of Massive Astrophysical Compact Halo Objects (MACHOs) by
observational searches for microlensing events in the halo of the
Milky Way and its immediate neighbors.

We developed a heuristic for detecting regions within data sets of
simulations of extreme astrophysical phenomena that are gravitationally
unstable. In this investigation, it was revealed that the marginally
stable objects that collapse into DBHs exhibit two qualitatively different
collapse scenarios depending on the mass of the DBH progenitor. We
termed DBHs whose progenitor fell into the more massive scenario Type
I and those whose progenitor fell into the low mass scenario Type
II. Type I DBH progenitors exhibit vanishing pressure on their surroundings
before collapse, and therefore can be treated fully by our adaptation
of the Chandrasekhar stability criterion. Type II DBH progenitors
exhibit finite pressure on their surroundings before collapse, and
thus cannot be fully treated by the Chandrasekhar stability criterion.
At the present time we can merely rule out the presence of Type II
DBH progenitors in a given, hypothetical data set only insofar as
no data points exhibit densities as high as the central density of
the least extreme DBH progenitor as shown in Figure \ref{fig:Peakiness}.
Due to the extremely small length scale of the core of a Type II DBH
progenitor, however, we should not consider the lack of evidence for
Type II DBHs in a data set to be completely conclusive. Sub-mesh scale
dynamics could in theory create the very small but very dense regions
that are necessary for Type II DBH progenitors but which are invisible
in the data set.

We analyzed a data set from a Type II supernova simulation conducted
by Christopher L. Fryer of Los Alamos National Laboratory and Patrick
A. Young of Arizona State University for the presence of DBH progenitors.
We found local density concentrations within a factor of 3-4 of both
the minimum mass and density necessary to create Type I DBHs. We did
not find any evidence of the densities necessary to create Type II
DBHs, although the simulation had sufficient resolution to detect
even the largest, most massive, and therefore least dense progenitors
(if they were there) for only a few tens of kilometers around its
center of mass.

Being within an order of magnitude of the conditions necessary for
the creation of DBHs within one of the milder members of the menagerie
of extreme astrophysical phenomena leads us to publish this work and
seek data sets of simulations of any highly energetic astrophysical
phenomena.

\section{Acknowledgements}

This research was made possible in part by a grant from the Maine
Space Grant Consortium, two Frank H. Todd scholarships, and a Summer
Graduate Research Fellowship and University Graduate Research Assistantship
from the University of Maine. The authors would like to thank Chris
Fryer of Los Alamos National Laboratory and Patrick Young of Arizona
State University for their invaluable data set. Andrew Hayes would
like to thank my wife Kate and daughter Evangeline for my entire universe.

\end{document}